\documentclass[a4paper,twoside]{article}
\usepackage{etex}
\usepackage[T1]{fontenc}
\usepackage[utf8]{inputenc}
\usepackage[spanish,english]{babel}
\usepackage{graphicx}
\usepackage[noindentafter,pagestyles]{titlesec}
\usepackage{textcomp}
\usepackage{dcolumn}
\usepackage{amsmath}
\usepackage{txfonts}
\usepackage{fancyhdr}
\usepackage{threeparttable}
\usepackage{xspace}
\usepackage{pictex}
\usepackage{comment}
\usepackage{float}
\usepackage{subcaption}
\usepackage{verbatim}
\usepackage{xcolor}
\usepackage{hyperref}

\usepackage{csquotes}

\usepackage[backend=biber,sorting=none]{biblatex} % Numeric style for numbered references
\DeclareFieldFormat[article]{title}{\mkbibquote{#1}} % Put quotes around the title
\DeclareFieldFormat{journaltitle}{#1} % No italics for journal title

\DeclareFieldFormat[book]{title}{\mkbibquote{#1}} % Put quotes around the title for books
\DeclareFieldFormat{booktitle}{#1} % No italics for book title

\renewbibmacro*{volume+number+eid}{%
  \printfield{volume}% Print volume
  \setunit*{\addcomma\space}% Comma after volume
  \printfield{number}% Print number (optional)
}

\DeclareBibliographyDriver{article}{%
  \printnames{author}% Print author names
  \addcomma\space
  \printtext{(}% Add opening parenthesis before year
  \printfield{year}% Print year in parentheses
  \printtext{)}% Add closing parenthesis after year
  \addcomma\space
  \printfield{title}% Print title in quotes
  \addcomma\space
  \printfield{journaltitle}% Journal title
  \addcomma\space
  \printfield{volume}% Print volume with "Vol."
  \addcomma\space
  \printfield{number}% Print issue number with "No."
  \addcomma\space
  \printfield{pages}% Print page numbers
  \addcomma\space
  \printfield{year}% Print page numbers
  \finentry
}

\DeclareBibliographyDriver{book}{%
  \printnames{author}% Print author names
  \addcomma\space
  \printtext{(}% Add opening parenthesis before year
  \printfield{year}% Print year in parentheses
  \printtext{)}% Add closing parenthesis after year
  \addcomma\space
  \printfield{title}% Print title in quotes
    \addcomma\space
  \printfield{edition}% Print publisher
  \addcomma\space
  \printfield{journaltitle}% Print publisher
  \addcomma\space
  \printfield{volume}% Print volume with "Vol."
  \addcomma\space
  \printfield{number}% Print issue number with "No."
  \addcomma\space
  \printfield{pages}% Print page numbers
  \addcomma\space
  \printfield{year}% Print page numbers
  \finentry
}
  
\hypersetup{
    colorlinks=true,   % Set to true for colored links
    linkcolor=black,    % Color of internal links
    filecolor=blue,    % Color of file links
    urlcolor=blue,     % Color of external links (like email and ORCID)
    citecolor=black,   % Color of citation links (black)
    pdftitle={Your Title},
    pdfpagemode=FullScreen,
}
\newcommand{\bmp}[1]{\begin{minipage}{#1\columnwidth}}
\newcommand{\emp}{\end{minipage}}

\newcommand{\bea}{\begin{eqnarray}}
\newcommand{\eea}{\end{eqnarray}}
\newcommand{\be}{\begin{equation}}
\newcommand{\ee}{\end{equation}}

\titleformat{\section}{\large\bfseries}{\thesection.}{.3em}{}
\titlespacing*{\section}{\leftmargini}{*3}{*3}
\titleformat{\subsection}{\bfseries}{\thesubsection}{.3em}{}
\titlespacing*{\subsection}{0pt}{*3}{*3}
\makeatletter
\def\@maketitle{%
  \newpage
  \null
  \vskip 2em%
  \begin{center}%
  \let \footnote \thanks
    {\fontsize{18}{22}\fontseries{b}\selectfont \@title \par}%
    \vskip 1.5em%
    {\normalsize
      \lineskip .5em%
      \begin{tabular}[t]{c}%
\@author
      \end{tabular}\par}%
    \vskip 1em%
    {\large \@date}%
  \end{center}%
  \par
  \vskip 1.5em}
\renewenvironment{abstract}{%
\if@twocolumn
\section*{\abstractname}%
\else
\quotation
\noindent{\bfseries\large \abstractname\vspace*{.3ex}\par}
\fi}
{\if@twocolumn\else\endquotation\fi}
\makeatother
\fancypagestyle{plain}{%
\fancyhf{} 
\fancyhead[L]{\large\MakeUppercase{\hspace*{20mm}} }
\fancyfoot[L]{\small }%
} 

\begin{document}
%
%
%  LIST PAPER TITLE, AUTHORS AND CONTACT INFORMATION
%
%  Add $^\dag$ to identify the corresponding author
%
%
\title{Combustion Instabilities in Complex Chamber Geometries of Solid Propellant Rocket Motors}
\author{%
    \Large \textsc{Juan M. Tiz\'on} \\
    E-mail: \href{mailto:jm.tizon@upm.es}{jm.tizon@upm.es} \\
    ORCID-ID: \href{https://orcid.org/0000-0002-8687-6657}{https://orcid.org/0000-0002-8687-6657} \\
    \\
    \large AND \\
    \\
    \Large \textsc{Antoni M. Barredo} \\
    E-mail: \href{mailto:antoni.barredo.juan@alumnos.upm.es}{antoni.barredo.juan@alumnos.upm.es} \\
    ORCID-ID: \href{https://orcid.org/0009-0007-9854-6624}{https://orcid.org/0009-0007-9854-6624} \\
    \\
    \small DEPARTAMENTO DE MECÁNICA DE FLUIDOS Y PROPULSIÓN AEROESPACIAL, \\
    \small ESCUELA TÉCNICA SUPERIOR DE INGENIERÍA AERONÁUTICA Y DEL ESPACIO (ETSIAE), \\
    \small UNIVERSIDAD POLITÉCNICA DE MADRID (UPM), \\
    \small PZA. DEL CARDENAL CISNEROS 3, 28040 MADRID, SPAIN
}

\date{}
\maketitle
\begin{abstract}
\noindent High-frequency combustion instabilities can lead to significant fluctuations in chamber pressure, affecting the structural integrity and performance of solid rocket motors. Since these instabilities manifest as acoustic oscillations during combustion, a mathematical model has been developed to calculate the chamber modes, providing an accurate method for predicting the acoustic behaviour of the combustion chamber. A novel method for discretising the Laplacian operator is introduced, which allows the calculation of the acoustic modes of complex chamber geometries. This approach uses an efficient numerical algorithm designed for unstructured mesh configurations.  Several computational examples are presented to demonstrate the application of the model to complex geometries typical of solid rocket motors. In addition, an analytical procedure is developed in these examples to aid the design process, providing engineers with critical data to optimise the stability and performance of propulsion systems.
This research provides valuable insights into the analysis of combustion instabilities, with broad implications for the design and optimisation of propulsion systems in aerospace engineering.
\end{abstract}

\begin{table}[h!]
\section*{Nomenclature}
    \centering
    \begin{minipage}[t]{0.45\textwidth}
        \centering
        \begin{tabular}{p{2cm} |l}
        \hspace{-8pt}Symbols & \\
        $A$ & Transfer function coefficient  \\
        $A_b$ & Area of the burning surface [$m^2$]  \\
        $A_g$ & Area of the nozzle throat [$m^2$]  \\
        $a$ & Speed of sound [$m/s$]  \\
        $B$ & Transfer function coefficient  \\
        $c^*$ & Characteristic velocity [$m/s$]  \\
        $E_n$ & Modal space norm  \\
        $f$ & Inhomogeneous boundary term \\
        $h$ & Inhomogeneous term  \\
        $K_g$ & Klemmung  \\
        $k$ &  Complex wave number [1/s]  \\
        $k^*$ &   Characteristic wave number [1/m]  \\
        $L^*$ &   Characteristic chamber length [m]  \\
        $\Delta l$ &   Mesh contour  \\
        $M$ &   Mach number  \\
        $\Dot{m}$ &   Mass flow rate [kg/s]  \\
        $n_s$ &   Pyrolysis constant  \\
        $\hat{n}$ &   Normal vector  \\ % check
        $p_c$ &   Chamber pressure [$N/m^2$]  \\
        $R$ &   Gas constant of mixture [$\frac{J}{kg K}$]  \\
        $R_g$ &   Nozzle transfer function  \\
        $R_p$ &   Burning surface transfer function \\
        \end{tabular}
    \end{minipage}
    \hfill
    \begin{minipage}[t]{0.45\textwidth}
        \centering
        \begin{tabular}{p{2cm}|l}
        $\dot{r}_b$ &   Regression rate [$m/s$]  \\
        $T_c$ &    Chamber temperature [K] \\
        $u$ &   Fluid velocity [m/s]  \\
        \hspace{-8pt}Greek symbols & \\
        $\alpha$ &   Growth rate [1/s]  \\
        $\gamma$ &   Relation specific heat capacities  \\
        $\varepsilon_c$ &   Convergent area relation \\
        $\vartheta_c$ &   Volume of the cavity [$m^3$]  \\
        $\kappa$ &   Small amplitude  \\
        $\lambda$ &  Complex root  \\
        $\rho$ &   Density of the mixture [$\frac{kg}{m^3}$]  \\
        $\psi_N$ &   Unperturbed mode  \\
        $\psi_n$ &   Eigenfunction  \\
        $\Omega$ &   Dimensionless frequency  \\
        $\omega$ &   Angular frequency [rad/s]  \\
        \hspace{-8pt}Sub-indexes & \\
        $( \ )_b$ &   Burning surface  \\
        $( \ )_l$ &   Convergent region \\
        $( \ )_p$ &   Propellant \\
        $( \ )_{t}$ &   Throat \\
        $( \ )_{w}$ &   Wall \\
        \end{tabular}
    \end{minipage}
\end{table}

\section{Introduction}

Combustion instabilities in solid propellant rocket motors refer to oscillatory phenomena that arise during the combustion process, leading to fluctuations in chamber pressure and thrust \cite{Price1959, Culick1968}. These instabilities can manifest as low-frequency (bulk mode) or high-frequency (acoustic mode) oscillations, each posing significant risks to the structural integrity and performance of the rocket \cite{Yang1995}. Analyzing these instabilities is crucial for ensuring the reliability and safety of rocket operations, as they can cause mechanical vibrations, structural damage, and even catastrophic failure \cite{Huang2009}. Understanding and mitigating combustion instabilities are essential for the design and development of more efficient and robust propulsion systems. \\

\noindent In general, the chamber medium supports unsteady wave motions produced by combustion due to local fluctuations in pressure or temperature. This can be linked to a positive feedback mechanism where fluctuations propagate and are supported by combustor dynamics \cite{Poinsot2017}. When instability occurs, it can lead to catastrophic consequences, such as increased heat transfer through the burning surface, raising combustion rates \cite{sutton2017rocket}. This can boost mean pressure, thrust, and mass flux while reducing combustion time. To prevent these phenomena, it is necessary to develop an analysis and model theory that encompasses the main ideas mentioned. This includes studying combustor dynamics, the impact of combustion on primary variables, and damping effects \cite{Culick2006}.\\

\noindent From the experimental data gathered in the literature, it is observed that most of the combustion instabilities are motions of a self-excited dynamical system \cite{Culick2006}. Initially, the oscillations exhibit linear behavior. If their growth rate is positive, their amplitude increases exponentially, indicating that the source terms exciting the oscillations exceed the acoustic losses \cite{Poinsot2005}. However, unless a combustor explosion or blow-off occurs, a limit cycle is eventually reached, where nonlinear effects become crucial \cite{Yang1995}. During this phase, the amplitude remains constant, signifying a zero growth rate. This can happen due to increased acoustic losses or a phase shift between unsteady pressure and heat release.\\

\noindent The instabilities in solid rocket motors can arise either from vortex shedding phenomena, which occur primarily in relatively large segmented grains or grains with circular slots, or from the excitation of the combustion chamber's acoustic modes \cite{sutton2017rocket, Flandro1995}. The instabilities with acoustic origin are involved with the acoustic modes of the chamber (longitudinal, tangential and radial). The frequency values depend on the geometry of the cavity and the speed of sound \cite{Dowling2003}. The method to determine which modes are unstable relies on the balance of energy supplied by the exciting mechanisms and the energy extracted by the damping processes. \\

\noindent Due to the linear character of our study, it is known that a small disturbance of an initial value will grow exponentially in time as \(e^{\alpha_g}t\), where $\alpha_g>0$, called the growth constant. In combustors, due to the processes that cause the simultaneous growth and decay of the disturbances, the stability of the mode is determined by the net growth constant written as
\begin{equation}\label{growth_const}
    \alpha = \alpha_g - \alpha_d=(\alpha)_{combustion}+(\alpha)_{nozzle}+(\alpha)_{condensed}+(\alpha)_{structure}+...
\end{equation}

\noindent The latter determines the stability of the mode, depending on its sign. Instabilities in solid rocket motors can originate from various energy sources, including combustion processes, mean flow, and their combination \cite{EMELYANOV2017161}. Analyzing these energy sources and the mechanisms that transfer energy to unsteady motion is critical for developing a theory to predict instabilities. Since these instabilities manifest as periodic oscillations with time-dependent amplitudes, the primary goal is to reduce these amplitudes to acceptable levels. The mechanisms influencing these instabilities are greatly affected by the geometry and state of the reactants mixture, with pressure coupling, velocity coupling at the combustion surface, and residual combustion (often associated with metal fuel additives like aluminum particles) being principal mechanisms \cite{VO201812}. Among these, the dynamics of the combustion surface are considered dominant, causing most combustion instabilities in solid rockets. The chemical energy released from combustion is transformed into mechanical motions in the combustion products. Thus, the response of this region's dynamics depends on a frequency range where combustion processes amplify pressure disturbances.\\

\noindent Damping effects, primarily due to the nozzle and particles in the flux, play a significant role. The nozzle acts as a permeable surface that damps pressure fluctuations effectively, especially when the chamber's acoustic modes are longitudinal \cite{marble1977acoustic}. Damping by particles occurs only if they are present in the combustion products, with the intensity proportional to the size and mass fraction of the particles. \\

\noindent With this,  the main purpose of the study is established: to calculate the growth and decay constants for the modes corresponding to the classical acoustic resonances.

\section{Mathematical model}

This section will establish the model defined to analyse the instability of different combustion chambers. The combustion and combustor dynamics will be first studied separately and coupled later to define a stability parameter. Moreover,  different models of damping mechanisms will be set.\\

\noindent Due to the dimensions of the combustion chamber being higher than the characteristic thickness of the burning surface, unidimensional descriptions of the combustion processes will be used to characterize the sub-systems' response models. Even though particle distributions of the heterogeneous propellants impose tridimensional distributions of the variables, the scale disparity will prevail, promoting the average of the results as a good strategy.

\subsection{Combustion dynamics model}
As stated previously, the combustion dynamics model can be seen as a burning surface with a thin thickness and a dynamic defined by a response function that feeds the cavity. Hypothetically, if there is an increase in pressure nearby the burning surface, it will decrease the distance between the gas region and the surface. Consequently, it will increase the heat flux through the surface and its temperature, resulting in a higher decomposition rate of the surface and the temperature gradient in the interior face of the solid. \\

\noindent Considering an endothermic decomposition of the solid, an increase in the pyrolysis rate produces a rise in the energy used, reducing the energy of the zone. On the other hand, the temperature of the surface tends to diminish due to the greater heat flux through the interior of the solid. In a general sense, the previous analysis can be seen as a process that stabilises the initial perturbation. Nevertheless, the previous statements are sequential, where the stabilising effect emerges with a time delay, allowing the possibility of different pressure conditions when it acts.\\

\noindent The essence of the mechanism is the fluctuation of the mass rate that leaves the burning surface due to the disturbances of pressure. In order to represent the mechanism of energy transfer, the combustion dynamics are modelled as a response function, which is defined as the ratio of the fluctuation of mass flow rate that comes out of the combustion region to the imposed fluctuation of the pressure \cite{Culick2006}

\begin{equation}\label{response_function}
    R_p = \frac{m'/\langle m \rangle}{p'/\langle p\rangle}
\end{equation}

 \noindent where ( )$'$ means fluctuation and $\langle \ \rangle$ is an average value. If it is assumed that the fluctuations are steady sinusoidal oscillations, the pressure and mass flux can be written as
 
\begin{equation}\label{variables_response}
    \begin{split}
        \frac{m'}{\langle m \rangle} = \frac{\hat{m}}{\langle m \rangle}e^{-i\omega t}\\
        \frac{p'}{\langle p\rangle} = \frac{\hat{p}}{\langle p\rangle}e^{-i\omega t}
    \end{split}
\end{equation}

\noindent where $\hat{( \ )}$ represents the amplitude of the oscillation. Due to the oscillations of mass flux rate not being in phase, generally speaking, with the pressure oscillations, the response function is complex. The real part represents when two fluctuations are in phase, establishing the growth and decay of the oscillations. It is therefore necessary to calculate the response function. A first approximation can be made using a simplified model \cite{Culick2006} that assumes:
\begin{itemize}
    \item Quasi-steady behavior of all processes, except unsteady conductive heat transfer in the condensed phase.
    \item Homogeneous and constant material properties, non-reacting condensed phase.
    \item One-dimensional variations in space.
    \item Conversion of condensed material to gas phase at an infinitesimally thin interface
\end{itemize}
The following model is called QSHOD and only considers unsteady heat transfer in the condensed phase due to the principal cause of the dynamics being a result of the coupling pressure. Despite the response functions suggested in the literature, it was demonstrated that all models were dynamically identical \cite{Culick1968}. Apart from the deficiencies present in the model, it has been used so far as an effective instrument to describe the phenomenology, and as a substitute to elaborate more ambitious models.\\

\noindent The burning processes occur almost entirely within a thin region adjacent to the propellant surface. Therefore, solving a problem related to the combustion surface requires solving three regions: the solid phase, the gas phase, and the interface. The general problem consists on solving the temperature field of each phase separately, and matching the solutions at the interface region \cite{Tizon2018, Culick2006}. \\

\noindent The considered reference system moves with the average velocity of the burning surface $\langle \dot{r}_b\rangle$. Therefore, the coordinate that defines the position of the burning surface also defines the regression velocity of the burning surface, $\dot{r}_b = \langle \dot{r}_b \rangle - \Dot{x}_s$, being considered positive when the regression of the surface moves to the interior of the solid. Once the previous problem is derived, the transfer function can be expressed as 
\begin{equation}
    R_p=\frac{m'/\langle \Dot{m}\rangle}{p'/\langle p \rangle}=\frac{c_1+n_s(\lambda -1)}{\lambda + \frac{A}{\lambda} + c_2}
\end{equation}

\noindent For $\lambda = 0$, meaning lack of fluctuations, the response function is
\begin{equation}
    \left(\frac{m'/\langle\Dot{m}\rangle}{p'/\langle p \rangle}\right)_{\lambda = 1}=n
\end{equation}

\noindent which imposes a series of conditions for the expression, resulting in an adjustment of the constants. Therefore, the final expression of the response function is
\begin{equation}
     R_p = \frac{nAB+n_s(\lambda-1)}{\lambda+\frac{A}{\lambda}-(1+A)+AB}
\end{equation}

\noindent where $n$ is the exponent of the Vieille law. The parameters $A$ and $B$ are experimental coefficients that must be obtained for each propellant.

\begin{figure}[H]
    \includegraphics[width = 0.6\textwidth]{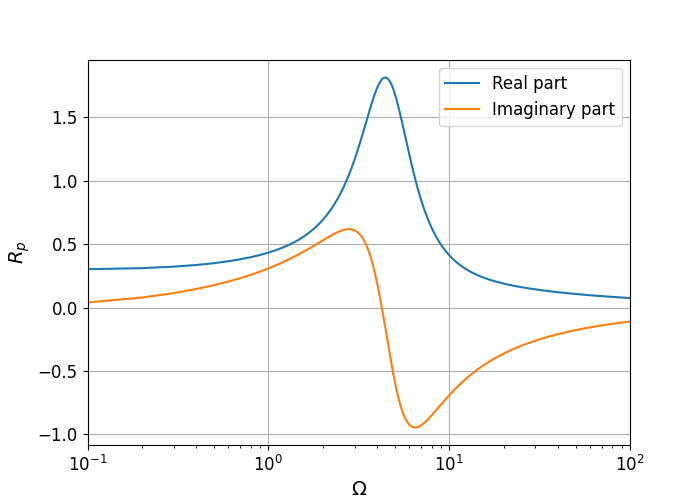}
    \centering
    \caption{Representation of the Real and Imaginary Parts of the QSHOD Response Function with $A=6$, $B=0.6$, $n=0.3$, $\gamma = 1.2$ and $n_s=0$.}
    \label{response_function_culick}
\end{figure}

\subsection{Combustor dynamics model}

 The purpose of this section is to establish a model of the combustor dynamics. The methods that will be used are based on modal expansion and space averaging. By using a two-parameter expansion of the equation of motion for reacting flows it is possible to end up with the following set of equations \cite{Culick2006}

\begin{equation}
    \nabla^2p'-\frac{1}{\langle a\rangle^2}\frac{\partial^2p'}{\partial t^2}=h
\end{equation}

% from the momentum equation
\noindent and the boundary condition for the pressure field being

\begin{equation}
    \hat{\textbf{n}}\cdot\nabla p'=-f
\end{equation}

\noindent For classical acoustics, the right hand side of the equations are

\begin{subequations}
    \begin{equation}
        h = \langle \rho\rangle\nabla \cdot \frac{1}{\rho}\mathcal{F'}-\frac{1}{\langle a\rangle^2}\frac{\partial \mathcal{P'}}{\partial t}
    \end{equation}
    \begin{equation}
        f=\langle \rho\rangle\frac{\partial \textbf{M}'}{\partial t}\cdot\hat{n}-\mathcal{F}'\cdot \hat{n}
    \end{equation}
\end{subequations}

\noindent Assuming the simplest model for the combustor dynamics as a single stationary wave, whose number of modes have been increased due to the combustion processes and nonlinear gas dynamical effects, a general solution to the inhomogeneous acoustic problem can be obtained by using  Green's function \cite{Culick2006}

\begin{equation}\label{pressure_2_green}
\begin{split}
    \hat{p}(\textbf{r})=     \psi_N(\textbf{r})\frac{\kappa}{{E_N}^2(k^2-{k_N}^2)}\left\{\iiint_V\psi_N(\textbf{r}_0)\hat{h}(\textbf{r}_0)dV_0+\oiint_S\psi_N(\textbf{r}_{0s})\hat{f}(\textbf{r}_{0s})dS_0\right\} \\
    +\kappa{\sum_{n=0}^\infty}^\prime\frac{\psi_n(\textbf{r})}{{E_n}^2(k^2-{k_n}^2)}\left\{\iiint_V\psi_n(\textbf{r}_0)\hat{h}(\textbf{r}_0)dV_0+\oiint_S\psi_n(\textbf{r}_{0s})\hat{f}(\textbf{r}_{0s})dS_0\right\}
\end{split}
\end{equation}

\noindent Equation \ref{pressure_2_green} is only consistent if the factor that multiplies the mode shape $\psi_N$ is equal to unity
\begin{equation}
    \frac{\kappa}{E_N^2(k^2-k_N^2)}\left\{\iiint_V\psi_N(\textbf{r}_0)\hat{h}(\textbf{r}_0)dV_0+\oiint_S\psi_N(\textbf{r}_{0s})\hat{f}(\textbf{r}_{0s})dS_0\right\} = 1
\end{equation}

\noindent From the rearrangement of the previous equation it is possible to obtain an expression for the complex wave number associated with the unperturbed mode $N$, if the inhomogeneous terms are known.
\begin{equation}\label{wavenumber_expression}
    k^2 = k_N^2 + \frac{\kappa}{E_N^2}\left\{\iiint_V\psi_N(\textbf{r}_0)\hat{h}(\textbf{r}_0)dV_0+\oiint_S\psi_N(\textbf{r}_{0s})\hat{f}(\textbf{r}_{0s})dS_0\right\}
\end{equation}

\noindent The imaginary part of the wavenumber determines the stability of the problem. \\

\subsubsection{Inhomogeneous boundary term} 

\noindent In this section a model of the inhomogeneous boundary term will be developed. Firstly, the different notations used during the section will be defined.
\begin{equation}\label{notation_unbound}
\begin{split}
    m'=\hat{m}e^{i\langle a\rangle kt} \\
    m'=\Tilde{m}e^{i\omega t}= \hat{m}e^{i\langle a\rangle k_{Re}t} \\
    \Tilde{m}=\hat{m}e^{i\langle a\rangle k_{Im}t}=\hat{m}e^{\alpha t} \\
    p'=\hat{p}e^{i\Bar{a}kt} \\
    \Tilde{p}=\hat{p}e^{i\langle a\rangle k_{Im}t}=\hat{p}e^{\alpha t} \\
\end{split}
\end{equation} 

\noindent while $\Tilde{m}$ contains the growth and decay factor of $m'$ in time, for $\hat{m}$ the factor is accounted for in the term \(e^{i\langle a\rangle k_{Im}t}\). Assuming \(u'=\hat{u}e^{i\langle a\rangle kt}\)
\begin{equation}\label{boundary_f}
    f=\Bar{\rho}\frac{\partial \Vec{u}'}{\partial t} \cdot \Vec{n} = \langle \rho\rangle i k \langle a\rangle(\Vec{\hat{M}}\cdot \Vec{n})e^{i\langle a\rangle kt}
\end{equation}

% M = u?
\noindent where $\Vec{n}$ is the normal vector outward to the surface. Recalling the expression of the response function defined in a previous section
\begin{equation}\label{response_newww}
    R_s(\Vec{r}_s,\omega) = \frac{\Dot{m}'/\langle\Dot{m}\rangle}{p'/\langle p\rangle} \quad \rightarrow \quad \frac{\Tilde{\Dot{m}}}{\langle \Dot{m}\rangle}=R_s\frac{\langle\Tilde{p}\rangle}{\langle p\rangle}
\end{equation}

\noindent It is important to remark that the response function $R_s(\Vec{r}_s,\omega)$ is defined as $R_p(\Vec{r}_s,\omega)$  in the region of the propellant, therefore, depends on the propellant parameters, the chemical reaction and temperatures. In a general sense, it is a function that depends on the position at the surface and $\omega$. If the properties of the propellant are assumed to be constant through the burning surface, the function will only depend on $\omega$. \\ 

\noindent As a result, it is useful to define a time lag for the mass flux rate and the pressure linked by the response function
\begin{equation}\label{time-lag_press_max}
    \begin{split}
        \Dot{m}'=\Tilde{\Dot{m}}e^{i\omega t} = \hat{\Dot{m}}e^{i\langle a\rangle k t} = ||\Tilde{\Dot{m}}||e^{i\varphi_m}e^{i\omega t}=||\hat{\Dot{m}}||e^{i\varphi_m}e^{i\langle a\rangle k t}\\
        p'=\Tilde{p}e^{i\omega t} = \hat{p}e^{i\langle a\rangle k t} = ||\Tilde{p}||e^{i\varphi_p}e^{i\omega t}=||\hat{p}||e^{i\varphi_p}e^{i\langle a\rangle k t}
    \end{split}
\end{equation}

\begin{equation}\label{pressure_response}
\begin{split}
    \langle \rho\rangle\Vec{n}\cdot \Vec{\hat{u}}(\Vec{r}_s,\omega)+\hat{\rho}(\Vec{r}_s,\omega)\Vec{n}\cdot \vec{\langle u\rangle} = -\hat{\Dot{m}}(\vec{r}_s,\omega)=\Tilde{\Dot{m}}e^{-i\langle a\rangle k_{Im}t}=
    \\
    =-\langle \Dot{m}\rangle R_s\frac{\hat{p}e^{i\langle a\rangle k_{Im}t}}{\langle p\rangle}e^{-i\langle a\rangle k_{Im}t}=-\langle \Dot{m}\rangle R_s(\vec{r}_s,\omega)\frac{\hat{p}}{\langle p\rangle}
\end{split}
\end{equation}

% Oscilaciones iséntropicas
\noindent where isentropic oscillations are assumed 
\begin{equation}\label{relation_pressure_velocity}
\begin{split}
    \hat{\rho}(\vec{r}_s,\omega)=\frac{\langle \rho\rangle\hat{p}}{\gamma \langle p\rangle}\\
    \vec{n}\cdot\vec{\langle u\rangle}=-\frac{\langle \Dot{m}\rangle}{\langle \rho\rangle}
\end{split}
\end{equation}

\noindent which gives
\begin{equation}\label{relation_velocityss}
    \langle \rho\rangle\vec{n}\cdot\vec{\hat{u}}(\vec{r}_s,\omega)=-\langle \Dot{m}\rangle R_s\frac{\hat{p}}{\langle p\rangle}+\frac{\langle \rho\rangle}{\gamma}\frac{\hat{p}}{\langle p\rangle}\frac{\langle \Dot{m}\rangle}{\langle \rho\rangle}=-\langle \Dot{m}\rangle\left(R_s-\frac{1}{\gamma}\right)\frac{\hat{p}}{\langle p\rangle}
\end{equation}

\noindent It can be observed that the factor $1/\gamma$ has a stabilizing effect, diminishing the real part of the response function. Defining \(R_s'=R_s-1/\gamma\) and introducing the Equation \ref{relation_velocityss} in the Equation \ref{boundary_f} gives
\begin{equation}\label{f_neew}
    f = \langle \rho\rangle\frac{\partial \Vec{u}'}{\partial t} \cdot \Vec{n}  = -k\frac{\langle \Dot{m}\rangle\langle a\rangle}{\langle p\rangle}[-\Im(R_s'(\vec{r}_s,\omega))+i\Re(R_s'(\vec{r}_s,\omega))]\hat{p}(\vec{r}_s,\omega)e^{i\langle a\rangle kt}
\end{equation}

\noindent Therefore, recalling the definition of $f$ depending on the factor $\kappa$
\begin{equation}
    f = \kappa \hat{f}e^{i\langle a\rangle kt}
\end{equation}

\noindent where 
\begin{equation}
    \hat{f} = -k\frac{\langle \Dot{m}\rangle\langle a\rangle}{\kappa\langle p\rangle}[-\Im(R_s'(\vec{r}_s,\omega))+i\Re(R_s'(\vec{r}_s,\omega))]\hat{p}(\vec{r}_s,\omega)
\end{equation} \\

\subsubsection{Analytical solution}

\noindent In this section the analytical solution for the wavenumber $k$ will be defined.  If the inhomogeneous term of the equation is not taken into account, \(\hat{h}(\vec{r})=0\), and is only considered as the inhomogeneous boundary term, the Equation \ref{wavenumber_expression} becomes
\begin{equation}\label{wavenumber_neew}
    k^2 = k_N^2 + \frac{\kappa}{E_N^2}\left\{\oiint_S\psi_N(\textbf{r}_{s})\hat{f}(\textbf{r}_{s})dS\right\}
\end{equation}

\noindent and introducing the previous expression of the inhomogeneous boundary term gives
\begin{equation}\label{wavenumber_final_2}
    k^2 = k_N^2 - \frac{\kappa}{E_N^2}\left\{\oiint_S\psi_N(\textbf{r}_{s})k\frac{\langle \Dot{m}\rangle\langle a\rangle}{\kappa\langle p\rangle}[-\Im(R_s'(\vec{r}_s,\omega))+i\Re(R_s'(\vec{r}_s,\omega))]\hat{p}(\vec{r}_s,\omega)dS\right\}
\end{equation}

\noindent If \(\hat{p}(\vec{r}_s,\omega)=\psi_N(\vec{r}_i)\) in first approximation, the previous equation can be expressed as
\begin{equation}\label{wavenumber_final}
    k^2 = k_N^2 - k\left\{\oiint_S\frac{\psi_N(\vec{r}_s)\psi_N(\vec{r}_s)}{E_N^2}\frac{\langle \Dot{m}\rangle\langle a\rangle}{\langle p\rangle}[-\Im(R_s'(\vec{r}_s,\omega))+i\Re(R_s'(\vec{r}_s,\omega))]dS\right\}
\end{equation}

\noindent where the value of $k$ can be obtained as a solution of a second-order complex equation. It has to be taken into account that the mean pressure and the mean speed of sound are considered homogeneous through the cavity. Rearranging the previous equation assuming \(k=k_{Re}+k_{Im}\) gives
\begin{equation}\label{decomp_wavenumber}
\begin{split}
        k_{Re}^2+k_{Im}^2+2ik_{Re}k_{Im}= k_{N}^2-(k_{Re}+ik_{Im})[-\mathcal{I}+i\mathcal{R}]=\\ =k_N^2+[k_{Re}\mathcal{I}+k_{Im}\mathcal{R}]-i[k_{Re}\mathcal{R}-k_{Im}\mathcal{I}]
\end{split}
\end{equation}

\noindent where
\begin{equation}\label{r_i_expressions}
    \begin{split}
        \mathcal{R} = \oiint_S\frac{\psi_N(\vec{r}_s)\psi_N(\vec{r}_s)}{E_N^2}\frac{\langle \Dot{m}\rangle\langle a\rangle}{\langle p\rangle}\Re(R_s'(\vec{r}_s,\omega))dS\\
        \mathcal{I} = \oiint_S\frac{\psi_N(\vec{r}_s)\psi_N(\vec{r}_s)}{E_N^2}\frac{\langle \Dot{m}\rangle\langle a\rangle}{\langle p\rangle}\Im(R_s'(\vec{r}_s,\omega))dS
    \end{split}
\end{equation}

\noindent which allows the separation of the equation into the real and imaginary parts.
\begin{equation}\label{k_sepparate}
    \begin{split}
        k_{Re}^2+k_{Im}^2=k_{N}^2+[k_{Re}\mathcal{I}+k_{Im}\mathcal{R}]\\
        2ik_{Re}k_{Im} = -[k_{Re}\mathcal{R}-k_{Im}\mathcal{I}]
    \end{split}
\end{equation}

\noindent Dividing the second equation by $k_{Re}$ gives
\begin{equation}\label{k_im_fnal}
    k_{Im} = -\frac{1}{2}[\mathcal{R}-\frac{k_{Im}}{k_{Re}}\mathcal{I}]
\end{equation}

\noindent Assuming that \(\frac{k_{Im}}{k_{Re}}\ll1\), the previous equation becomes
\begin{equation}
    k_{Im}\simeq-\frac{1}{2}\oiint_S\frac{\psi_N(\vec{r}_s)\psi_N(\vec{r}_s)}{E_N^2}\frac{\langle \Dot{m}\rangle\langle a\rangle}{\langle p\rangle}\Re(R_s'(\vec{r}_s,\omega))dS=-\frac{1}{2}\mathcal{R}
\end{equation}

\noindent As it is observed, the most relevant parameters for the stability of the system are \(\langle \Dot{m}\rangle, \ \langle p\rangle, \ \langle a\rangle,\) the response function $R_s'$(in general, dependent on the propellant) and the geometry of the cavity ($\psi_N$). Moreover, it can be checked that values of $\Re(R_s'(\vec{r}_s,\omega))>0$ become the values $k_{Im}<0$, which correspond to an increase of the instability of the system (\(k=\frac{1}{\langle a\rangle}(\omega-i\alpha), \ \alpha>0 \Rightarrow \) instability).\\

\noindent From a physical point of view, the previous conclusion can be interpreted as suggesting that the time-lag between the pressure and the mass flow rate ($\hat{p}$ and $\hat{m}$) is such that $\Re(R_s'(\vec{r}s,\omega))>0$, meaning that an increase in pressure leads to an increase in the burning rate and mass flow rate at the interfacial region. Therefore, it can be seen that the factor $\hat{p}$ directly influences the factor $\hat{m}$, inducing instability. Mathematically, the stability linked to $k_{Im}$ is linearly related to -$\mathcal{R}$.

\subsection{Nozzle damping mechanisms}

In this section, models of different nozzle damping mechanisms will be established, which provide negative contributions to the growth constant $\alpha_g$, stabilizing the oscillations. \\

\subsubsection{First approximation} 

\noindent The response function of the nozzle can be defined as $R_s'(\vec{r}_g,\omega)=R_g(\vec{r}_g,\omega)$. Assuming that the nozzle throat is choked ($M_g=1$), the mass flow parameter will be constant
\begin{equation}\label{gasto}
    \frac{\dot{m}\sqrt{RT_t}}{p_tA_g}=\sqrt{\gamma}\Gamma(\gamma)
\end{equation}

\noindent Nevertheless, the values of primary variables will be perturbed as
\begin{equation}
    T_t=\langle T_t\rangle+T_t' \ ; \quad p_t=\langle p_t\rangle+p_t' \ ; \quad \dot{m} =\langle \dot{m}\rangle + \dot{m}'
\end{equation}

\noindent Expanding the previous expressions results in
\begin{subequations}
\begin{equation}
    \sqrt{T_t} = \sqrt{\langle T_t\rangle}+\frac{T_t'}{2\sqrt{\langle T_t\rangle}}+\mathcal{O}(T_t'^2)\simeq\sqrt{\langle T_t\rangle}
\end{equation}
\begin{equation}\label{pressure_eq_expand}
\begin{split}
    p_t=(\langle p\rangle+p')\left(1+\frac{\gamma-1}{2}M_g^2\right)^{\frac{\gamma}{\gamma-1}}=(\langle p\rangle+p')\left(1+\frac{\gamma-1}{2}(1+M_g'^2)\right)^{\frac{\gamma}{\gamma-1}}=\\
    =\langle p\rangle\left(1+\frac{\gamma-1}{2}(1+M_g'^2)\right)^{\frac{\gamma}{\gamma-1}}+p'\left(1+\frac{\gamma-1}{2}(1+M_g'^2)\right)^{\frac{\gamma}{\gamma-1}}
\end{split}
\end{equation}
\end{subequations}

\noindent Assuming $M_g'\ll1$, the Equation \ref{pressure_eq_expand} becomes
\begin{equation}
\begin{split}
    p_t=(\langle p_t\rangle +p'_t)=(\langle p\rangle+p')\left(1+\frac{\gamma-1}{2}\right)^{\frac{\gamma}{\gamma-1}}\\
    \langle p_t\rangle=\langle p\rangle\left(1+\frac{\gamma-1}{2}\right)^{\frac{\gamma}{\gamma-1}} \quad ; \quad p_t'=p'\left(1+\frac{\gamma-1}{2}\right)^{\frac{\gamma}{\gamma-1}}
\end{split}
\end{equation}

\noindent Introducing the previous expressions in the perturbed expression of the mass flow rate gives
\begin{equation}
\begin{split}
    \langle \dot{m}\rangle+\dot{m}'=\frac{\sqrt{\gamma}\Gamma(\gamma)A_g}{\sqrt{R\langle T_t\rangle}}(\langle p_t\rangle+p_t')\\
    \langle \dot{m}\rangle=\frac{\sqrt{\gamma}\Gamma(\gamma)A_g}{\sqrt{R\langle T_t\rangle}}\langle p_t\rangle \quad ; \quad     \dot{m}'=\frac{\sqrt{\gamma}\Gamma(\gamma)A_g}{\sqrt{R\langle T_t\rangle}}p_t'
\end{split}
\end{equation}

\noindent Dividing the expression for mass flow fluctuations  by the expression of the mean mass flow gives
\begin{equation}
\begin{split}
    \frac{\dot{m}'}{\langle \dot{m}\rangle}=\frac{p_t'}{\langle p_t\rangle}\\
\end{split}
\end{equation}

\noindent Expanding the previous result in harmonics, the response function can be obtained
\begin{equation}
    \begin{split}
        \frac{\hat{\Dot{m}}}{\langle \Dot{m}\rangle}e^{i\langle a\rangle kt}=\frac{\hat{p}_t}{\langle p_t\rangle}e^{i\langle a\rangle kt}\\
        \frac{\hat{\Dot{m}}}{\langle \Dot{m}\rangle}=R_g\frac{\hat{p}_t}{\langle p_t\rangle} \Rightarrow R_g = 1
    \end{split}
\end{equation}

\noindent which means that the pressure fluctuations produce the same fluctuation as, and in phase with, the mass flow. Therefore, $R_s'(\vec{r}_g,\omega)=R_g(\vec{r}_g,\omega)=1$. \\

\subsubsection{Tsien model}

\noindent The previous statement is a first approximation of the presence of the nozzle. In this section, a more elaborated model \cite{TSIEN1952} will be developed, which will allow the implementation of the influence of different nozzle conditions. Therefore, the main purpose will be to compute the ratio \((\rho'/\rho+u'/u)/(p'/p)\) at the entrance of the nozzle as a function of the oscillation frequency $\omega$. \\

\noindent An uni-dimensional nozzle will be assumed, where each section has uniform conditions. The conditions at the entrance of the nozzle are fixed by assuming that the chamber temperature is not changed by variations in pressure. Due to the flow being supersonic in the divergence of the nozzle, it can be assumed that the propagation of oscillations must be toward the downstream direction.\\

\noindent The formulation of the problem is based on defining three equations for the three dependent variables \((\rho'/\rho\), \(p'/p\) and \(u'/u\)). These are the continuity, dynamic, and entropy conservation equations, where the first-order fluctuations terms have been retained
\begin{subequations}
\begin{equation}
    \frac{\partial }{\partial t}\left(\frac{\rho'}{\rho}\right)+u\frac{\partial}{\partial x}\left(\frac{\rho'}{\rho}+\frac{u'}{u}\right)=0
\end{equation}
\begin{equation}
    \frac{\partial}{\partial t}\left(\frac{u'}{u}\right)+\left(\frac{\rho'}{\rho}+2\frac{u'}{u}\right)\frac{du}{dx}+u\frac{\partial}{\partial x}\left(\frac{u'}{u}\right)=\left(\frac{p'}{p}\right)\frac{du}{dx}-\frac{p}{\rho u}\frac{\partial}{\partial x}\left(\frac{p'}{p}\right)
\end{equation}
\begin{equation}
    \left(\frac{\partial}{\partial t}+u\frac{\partial}{\partial x}\right)\left[\left(\frac{p'}{p}\right)-\gamma\left(\frac{\rho'}{\rho}\right)\right]
\end{equation}
\end{subequations}

\noindent The Tsien model offers a different approach to the response function, depending on the magnitude of the frequencies \cite{TSIEN1952}.
\begin{align}
    R_g(\beta) = 1+i\beta\left\{\frac{\gamma+1}{4\gamma}\frac{log(\frac{1}{z_l})}{1-z_l}-\frac{\gamma-1}{4\gamma}\right\} \quad\quad \beta\ll 1 \\
    R_g(\beta)=1+\frac{1}{\gamma M_l}  \quad \quad \beta \gg 1
\end{align}

\noindent where 
\begin{equation}
    z_l = \frac{\frac{\gamma+1}{2}M_l^2}{1+\frac{\gamma-1}{2}M_l^2} \quad \quad \beta=\frac{\omega}{u_l}x_l
\end{equation}

\noindent being $M_l$ the Mach number at the nozzle entrance, $u_l$ the entrance velocity of the nozzle and $x_l$ the $x$-coordinate at the entrance of the nozzle, where the $x$-axis origin has to be treated adequately. It can be observed that when $\beta$ is not exactly zero, the transfer functions have a small 'lead component' proportional to the frequency. Moreover, it can be appreciated from the previous equations, that the first approximation made previously fits the response function of the nozzle for frequencies near to zero. For large frequencies, it is observed that the transfer function is real. Thus, the mass flux rate and the pressure are in phase. \\

\subsubsection{Different proposal} 

\noindent A different approach to the nozzle transfer function has been proposed, inspired by the Tsien model. The main goal of the proposal is to establish a correspondence between the frequencies contained in the range constrained by small dimensionless frequencies $\Omega\ll1$ and large dimensionless frequencies $\Omega\gg1$. A dimensionless frequency limit value is defined, $\Omega_c$ , which will serve as a parameter and whose influence on the stability value will be studied. Therefore, in order to set the evolution of the damping of the nozzle, the following transfer function is established
\begin{align}
    R_g(\Omega) = 1+ia_x\frac{\Omega}{\Omega_c} \quad\quad \Omega_c > \Omega\\
    \label{ref2}
    R_g(\Omega)=1+\frac{1}{\gamma M_c}  \quad \quad \Omega_c \leq \Omega
\end{align}

\noindent where the value of $a_x$ is obtained by equating the module of the transfer function when $\Omega = \Omega_c$ to the Equation \ref{ref2}.

\subsection{Coupled model}

Once the different parts of the system have been modelled, the integral of the Equation \ref{wavenumber_final} will be evaluated at the different boundaries of the system, in order to obtain the stability parameter. Recalling Equation \ref{wavenumber_final}
\begin{equation}\label{wavenumber_final_3}
    k^2 = k_N^2 - k\left\{\oiint_S\frac{\psi_N(\vec{r}_s)\psi_N(\vec{r}_s)}{E_N^2}\frac{\langle \Dot{m}\rangle\langle a\rangle}{\langle p\rangle}[-\Im(R_s'(\vec{r}_s,\omega))+i\Re(R_s'(\vec{r}_s,\omega))]dS\right\}
\end{equation}

\noindent The integral has to be evaluated at the different surface boundaries. Moreover, assuming that the properties are constant, the response functions do not depend on the position and can be extracted from the integrals.
\begin{equation}\label{k_expression}
    k^2 = k_N^2 - k\left\{\mathcal{P} + \mathcal{G}\right\}
\end{equation}

\noindent where
\begin{equation}
\begin{split}
    \mathcal{P}= i\frac{\langle \Dot{m}_b\rangle\langle a\rangle}{\langle p\rangle}R_p'(\omega)\iint_P\frac{\psi_N(\vec{r}_p)\psi_N(\vec{r}_p)}{E_N^2}dS_p\\
    \mathcal{G}= i\frac{\langle \Dot{m}_l\rangle\langle a\rangle}{\langle p\rangle}R_g(\omega)\iint_l\frac{\psi_N(\vec{r}_l)\psi_N(\vec{r}_l)}{E_N^2}dS_l\\
\end{split}
\end{equation}

\noindent where the subscript $l$ indicates the entrance of the convergent of the nozzle. Announcing the following expressions
\begin{equation}
    \begin{split}
        \dot{m} = \langle \Dot{m}_b\rangle A_b=\langle \Dot{m}_g\rangle A_g=\langle \Dot{m}_l\rangle A_l; \quad \dot{m} = \frac{p_cA_g}{c^*}\\
        K_g = \frac{A_b}{A_g}; \quad \varepsilon_c = \frac{A_c}{A_g}; \quad L^*=\frac{\vartheta_c}{A_g}; \quad \frac{\langle a\rangle}{c^*}=\sqrt{\gamma}\Gamma(\gamma); \quad \langle p\rangle=p_c
    \end{split}
\end{equation}

\noindent and introducing it in the Equation \ref{k_expression} gives
\begin{subequations}\label{k_expressiong}
\begin{equation}\label{k_expressions1}
    k^2 = k_N^2 - k\sqrt{\gamma}\Gamma(\gamma)\left\{i\frac{R_p'(\omega)}{K_g}\iint_P\frac{\psi_N(\vec{r}_p)\psi_N(\vec{r}_p)}{E_N^2}dS_p - i\frac{R_g(\omega)}{\varepsilon_c}\iint_l\frac{\psi_N(\vec{r}_l)\psi_N(\vec{r}_l)}{E_N^2}dS_l\right\}
\end{equation}
\begin{equation}\label{k_expressions2}
    k^2 = k_N^2 - k\frac{\sqrt{\gamma}\Gamma(\gamma)}{L^*}\left\{iR_p'(\omega)\frac{\iint_p\psi_N(\vec{r}_p)\psi_N(\vec{r}_p)\frac{dS_p}{A_b}}{\iiint_V\psi_N(\vec{r})\psi_N(\vec{r})\frac{dV}{\vartheta_c}} - \frac{iR_g(\omega)}{A_c/A_b}\frac{\iint_g\psi_N(\vec{r}_g)\psi_N(\vec{r}_g)\frac{dS_g}{A_b}}{\iiint_V\psi_N(\vec{r})\psi_N(\vec{r})\frac{dV}{\vartheta_c}}\right\}
\end{equation}
\end{subequations}

\noindent where it is seen that with dimensionless integrals the wavenumber depends on $L^*$ and, with dimension integrals, it depends on the Klemmung value and the relation of areas of the convergent region of the nozzle. It is to be noted that the contribution of the nozzle subtracts the oscillations due to its dampness. Defining the pulse as
\begin{equation}
    \Omega = \frac{\alpha_p\omega}{\Dot{r}_b^2}
\end{equation}

\noindent and introducing it in the expression of the wavenumber, gives
\begin{equation}\label{wavenumber_nice}
    k=\frac{1}{a}(\omega-i\alpha) \rightarrow k\frac{ a\alpha_p}{\Dot{r}_b^2}=\frac{\alpha_p}{\Dot{r}_b^2}(\omega-i\alpha)=(\Omega-i\Sigma)
\end{equation}

\noindent Defining the wavenumber $k^*$, with the properties of the propellant as
\begin{equation}
    k^*=\frac{\Dot{r}_b^2}{c^*\alpha_p}
\end{equation}

\noindent where $\Dot{r}_b$ is the regression rate and $\alpha_p$ is the thermal difussivity of the propellant. For as input data such as $\Dot{r}_b\sim0.01m/s, \ \alpha_p\sim10^{-7}m^2/s, \ c^*\sim1000m/s$, the value of the characteristic wavenumber is $k^*\sim1m^{-1}$

\noindent Assuming $\langle a\rangle \simeq c^*$ and dividing the Equation \ref{wavenumber_nice} by the previous definition gives the stability value
\begin{equation}
    \frac{k}{k^*}=(\Omega-i\Sigma)
\end{equation}

\noindent Introducing the stability value in the Equations \ref{k_expressiong} gives
\begin{subequations}\label{k_entrekstar}
\begin{equation}
    \left(\frac{k}{k^*}\right)^2 = \left(\frac{k_N}{k^*}\right)^2 - \left(\frac{k}{k^*}\right)i\frac{\sqrt{\gamma}\Gamma(\gamma)}{k^*}\left\{\frac{R_p'(\omega)}{K_g}\mathcal{J}_p - \frac{R_g(\omega)}{\varepsilon_c}\mathcal{J}_g\right\}
\end{equation}
\begin{equation}
    \left(\frac{k}{k^*}\right)^2 = \left(\frac{k_N}{k^*}\right)^2 - \left(\frac{k}{k^*}\right)i\frac{\sqrt{\gamma}\Gamma(\gamma)}{k^*L^*}\left\{R_p'(\omega)\mathcal{\Tilde{J}}_p - \frac{R_g(\omega)}{\varepsilon_c/K_g}\mathcal{\Tilde{J}}_g\right\}
\end{equation}
\end{subequations}

\noindent where 
\begin{equation}
    \begin{split}
    \mathcal{J}_p=\frac{\iint_p\psi_N(\vec{r}_p)\psi_N(\vec{r}_p)dS_p}{\iiint_V\psi_N(\vec{r})\psi_N(\vec{r})dV}; \quad     \mathcal{\Tilde{J}}_p=\frac{\iint_p\psi_N(\vec{r}_p)\psi_N(\vec{r}_p)\frac{dS_p}{A_b}}{\iiint_V\psi_N(\vec{r})\psi_N(\vec{r})\frac{dV}{\vartheta_c}}; \quad \mathcal{J}_p=\frac{K_g}{L^*}\mathcal{\Tilde{J}}_p\\
    \mathcal{J}_g=\frac{\iint_g\psi_N(\vec{r}_g)\psi_N(\vec{r}_g)dS_g}{\iiint_V\psi_N(\vec{r})\psi_N(\vec{r})dV}; \quad     \mathcal{\Tilde{J}}_g=\frac{\iint_g\psi_N(\vec{r}_g)\psi_N(\vec{r}_g)\frac{dS_g}{A_b}}{\iiint_V\psi_N(\vec{r})\psi_N(\vec{r})\frac{dV}{\vartheta_c}}; \quad \mathcal{J}_g=\frac{K_g}{L^*}\mathcal{\Tilde{J}}_g
    \end{split}
\end{equation}

\noindent It is apparent that the Equations \ref{k_entrekstar} are complex second order equations with $k/k^*$ as solutions. Moreover, it only depends on the integral factors whose value is related to the geometry of the cavity (mode shapes), the relation of the specific heats $\gamma$, the transfer functions of the propellant and the nozzle, $K_g$ and $k^*$. Therefore, the stability of the system depends on the geometry of the chamber, the nozzle, and the propellant.

\section{Laplacian discretization}

The different cavity geometries will be discretized in an unstructured triangular mesh. Therefore, to compute the eigenvalues and eigenvectors of the homogeneous wave equation problem in a $\Omega$ domain
\begin{subequations}
\begin{equation}\label{problem_laplace}
\nabla^{2}p'+k^{2}p'=0 \quad  \vec{x}\in \Omega
\end{equation}
\begin{equation}\label{problema_laplace_bc}
\nabla p'\cdot n=0 \quad     \vec{x}\in \partial\Omega
\end{equation}
\end{subequations}

\noindent it is necessary to make a discretization of the Laplacian operator.  \\

\noindent Firstly, the basic structure of the mesh and its naming will be described. It consists of a node-centered scheme $i$ which belongs to $N$ triangles. The minimal control volume is defined by the center node $i$ connected to the node $j$ through the edge $il$ which at the same time belongs to both triangles, $j$ and $j+1$. Finally, the triangle $j$ has a third node named $j-1$, and the triangle $j+1$ has a node $j+1$. In addition, it is necessary to describe the contours $(\Delta l)_{j-}$ y $(\Delta l)_{j+}$ that connect the centroid of the triangle $j$ to the middle point of each edge connected to the center node $i$. It has to be noted that the same index is used to identify the triangles and the outer nodes for the purpose of a better future rearrangement of the equations.

\begin{figure}[H]
    \begin{subfigure}{0.5\textwidth}
        \includegraphics[height=5cm]{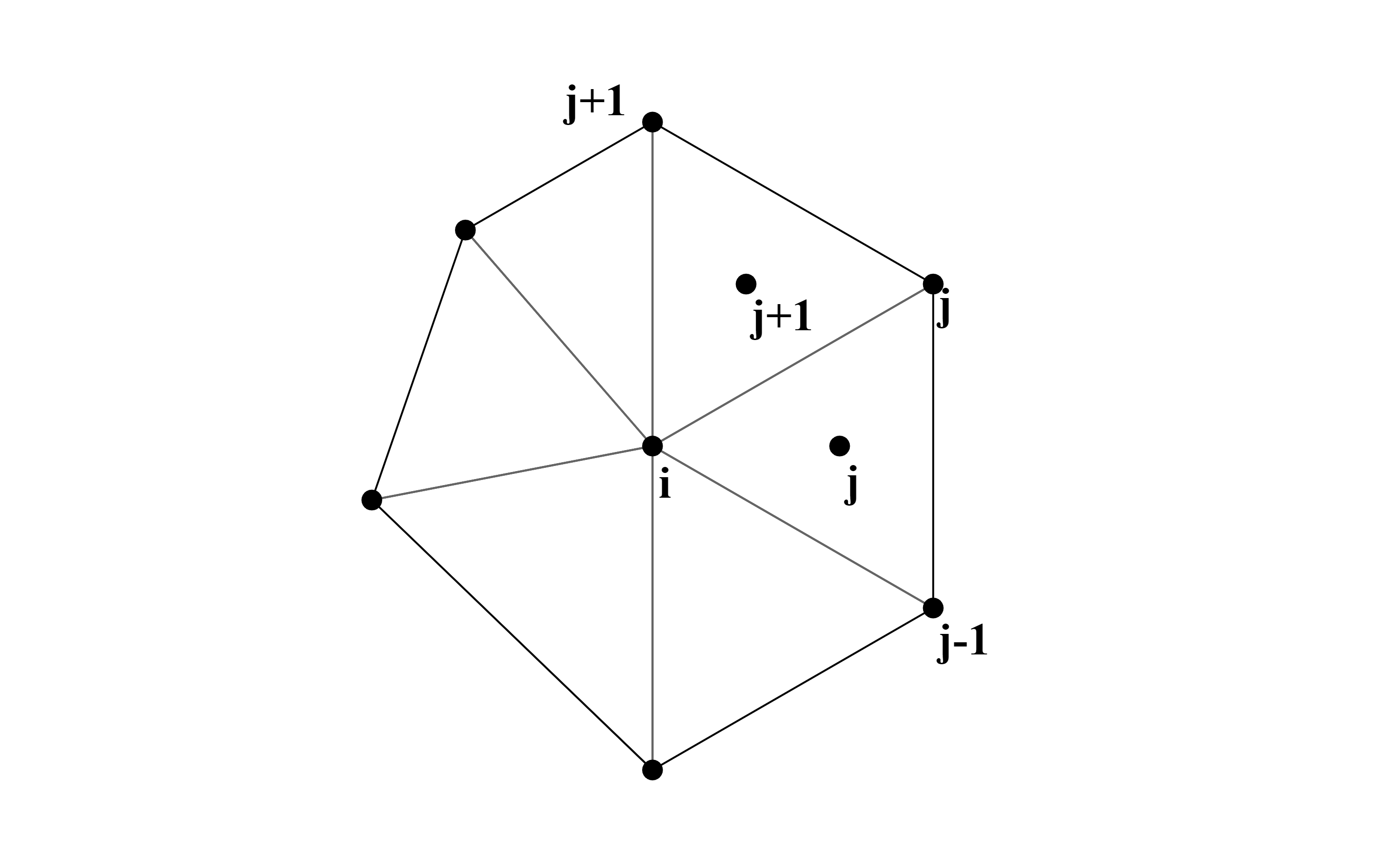}
        \centering
        \caption{Basic structure.}
        \label{basic_str}
    \end{subfigure}
    \begin{subfigure}{0.5\textwidth}
        \includegraphics[height=5cm]{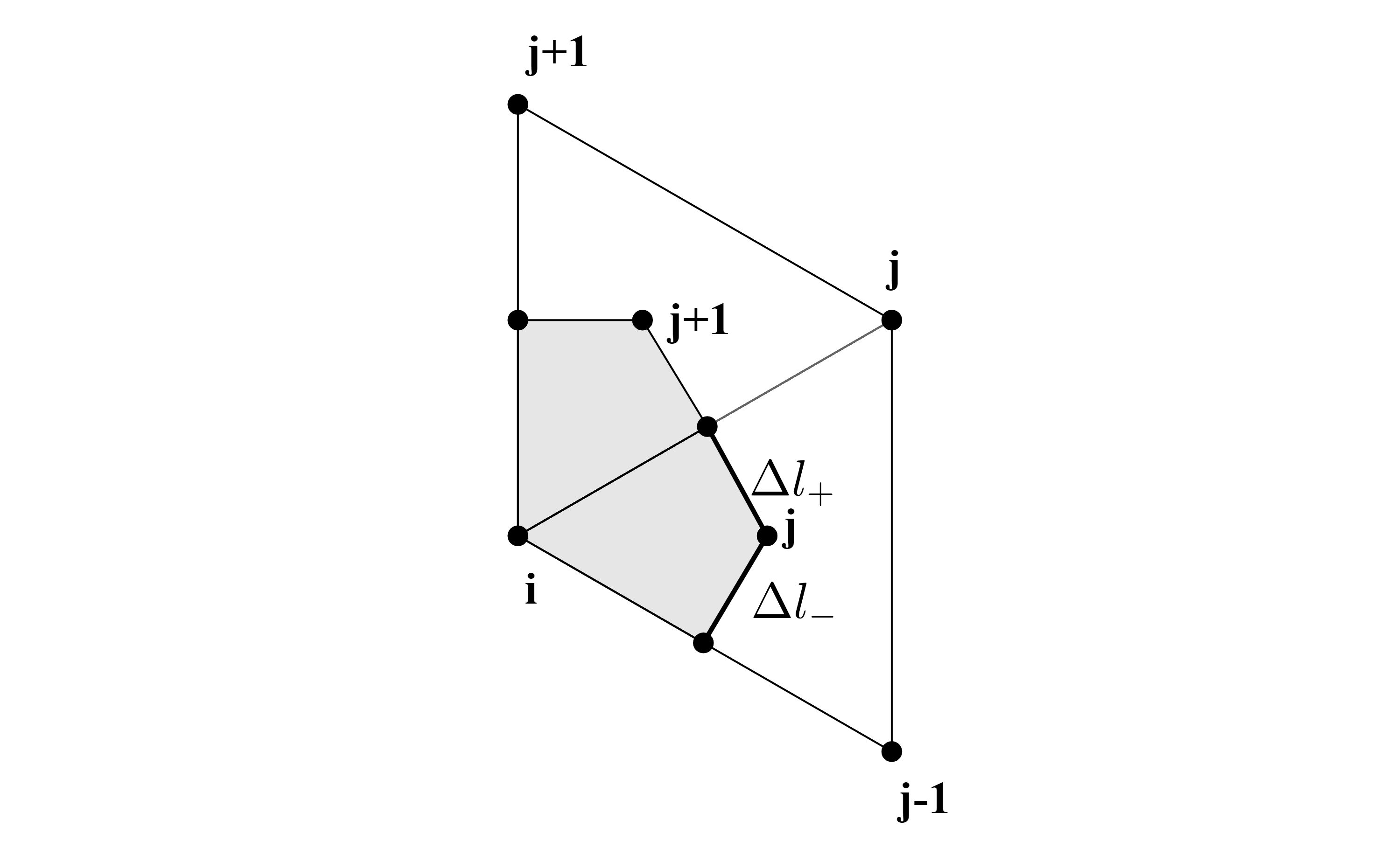}
        \centering
        \caption{Minimal control volume.}
        \label{naextrf2}
    \end{subfigure}
    \caption{Unstructured triangular mesh scheme.}
    \label{mod_rect_n2m3}
\end{figure}

\noindent Announcing the divergence theorem in the form
\begin{equation}\label{teo_divergen}
    \iint_{\Omega} \nabla^{2}p' d\Omega = \oint_{\partial\Omega} \nabla p'\cdot n \ dl
\end{equation}

\noindent which in a discretized form can be expressed as
\begin{equation}\label{teo_divergen_dicret}
    (\nabla^{2}p')_{i}\sum_{j=1}^{N}\frac{1}{3}A_{j}=\sum_{j=1}^{N}\vec{U}_{j}\cdot(\vec{n}_{j+}(\Delta l)_{j+}+\vec{n}_{j-}(\Delta l)_{j-})
\end{equation}

\noindent where \(\vec{U}_{j}=(\nabla p')_{j}\) is the discretized gradient of $p'$ in the triangle $j$. \\

\noindent Using this formulation, it is straightforward to apply the Newman boundary condition ignoring the integral on the appropriate sides. This formulation already limits the contribution of the sides of the boundary. However, another condition must be applied ad hoc.

\begin{figure}[H]
    \includegraphics[height=6cm]{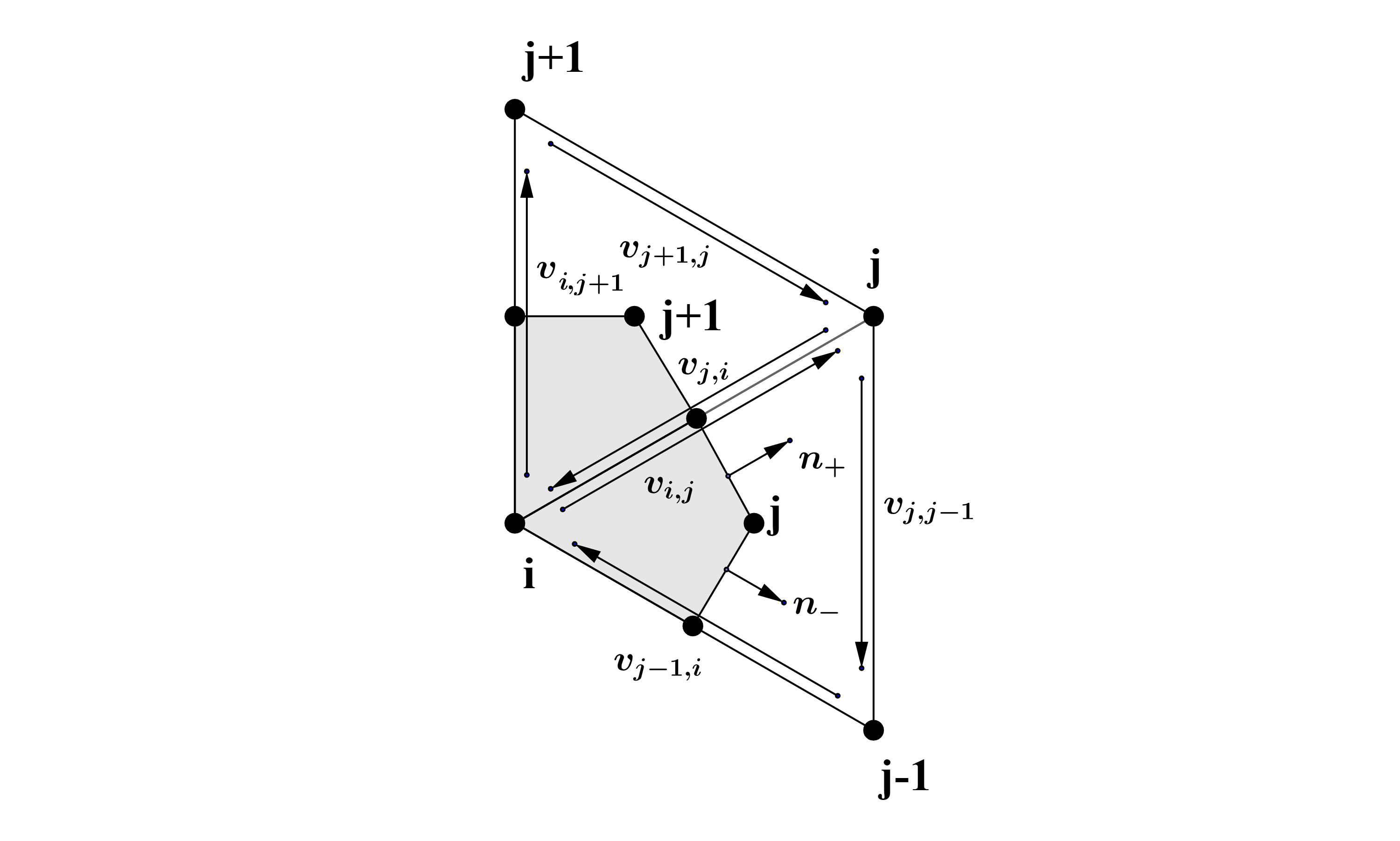}
    \centering
    \caption{Vectors definition of the minimal volume control.}
    \label{vectors_mesh}
\end{figure}

\noindent Firstly, the outward normal of the triangle $j$ will be defined. Therefore, the following vectors are established
\begin{equation}\label{vec1}
\begin{split}
    \vec{v}_{j,j-1}=(x_{j-1}-x_{j})\vec{i}+(y_{j-1}-y_{j})\vec{j}\\
    \vec{v}_{j-1,i}=(x_{i}-x_{j-1})\vec{i}+(y_{i}-y_{j-1})\vec{j}\\
    \vec{v}_{i,j}=(x_{j}-x_{i})\vec{i}+(y_{j}-y_{i})\vec{j}\\
\end{split}
\end{equation}

\noindent The outward normals of the triangle $j$ are defined as 
\begin{subequations}\label{normales}
\begin{equation}\label{normal_neg}
    \vec{n}_{j-}(\Delta l)_{j-}=\frac{1}{3}\left\{\left[y_{j}-\frac{1}{2}(y_{i}+y_{j-1})\right]\vec{i}+\left[x_{j}-\frac{1}{2}(x_{i}+x_{j-1})\right]\vec{j}\right\}
\end{equation}
\begin{equation}\label{normal_pos}
    \vec{n}_{j+}(\Delta l)_{j+}=\frac{1}{3}\left\{-\left[y_{j-1}-\frac{1}{2}(y_{i}+y_{j})\right]\vec{i}+\left[x_{j-1}-\frac{1}{2}(x_{i}+x_{j})\right]\vec{j}\right\}
\end{equation}
\end{subequations}

\noindent Replacing the expressions \ref{normales} in the Equation \ref{teo_divergen_dicret} gives
\begin{equation}\label{laplace_discrete_norm}
    (\nabla^{2}p')_{i}\sum_{j=1}^{N}\frac{1}{3}A_{j}=\frac{1}{3}\sum_{j=1}^{N}\left[-U_{j,x}\frac{3}{2}(y_{j-1}-y_{j})+U_{j,y}\frac{3}{2}(x_{j-1}-x_{j})\right]
\end{equation}

\noindent For the axisymmetric case, the integral must be correctly evaluated
\begin{equation}\label{laplace_discrete_norm_axil}
    (\nabla^{2}p')_{i}\sum_{j=1}^{N'}y_{i}\frac{1}{3}A_{j}=\frac{1}{3}\sum_{j=1}^{N}y_{j}\left[-U_{j,x}(y_{j-1}-y_{j})+U_{j,y}(x_{j-1}-x_{j})\right]
\end{equation}

\noindent The integral of the right side is not evaluated correctly. It should be the radius at the middle point of each segment instead of the radius at the centroid of each triangle. Nevertheless, this simplification provides an expression that is, in a general sense, similar to the bidimensional expression. The axisymmetric form will be preserved for the following procedures, knowing that the bidimensional expression is obtained by dropping the factors $y_{i}$ and $y_{j}$.\\

\noindent On the other hand, the components of the gradient of the triangle $j$ are defined as
\begin{subequations}\label{gradient_components}
\begin{equation}
    U_{jx}=\frac{1}{2A_{j}}[p'_{i}(y_{j-1}-y_{j})+p'_{j-1}(y_{j}-y_{j})+p'_{j}(y_{i}-y_{j-1})]
\end{equation}
\begin{equation}
    U_{jy}=\frac{1}{2A_{j}}[-p'_{i}(x_{j-1}-x_{j})-p'_{j-1}(x_{j}-x_{j})-p'_{j}(x_{i}-x_{j-1})]
\end{equation}
\end{subequations}

\noindent obtained from the equation of the plane
\begin{equation}
    \begin{vmatrix}
    x-x_{i} & y-y_{i} & p'-p'_{i}\\
    x_{j-1}-x_{i} & y_{j-1}-y_{i} & p'_{j-1}-p'_{i}\\
    x_{j}-x_{i} & y_{j}-y_{i} & p'_{j}-p'_{i}
    \end{vmatrix}
    = 0
\end{equation}

\noindent As can be seen, that the gradient is constant in each triangle.
\noindent Substituting \ref{gradient_components} in the Equation \ref{laplace_discrete_norm_axil} gives
\begin{equation}\label{eq_grad}
    (\nabla^{2}p')_{i}\sum_{j=1}^{N}y_{i}\frac{1}{3}=\frac{1}{4}\sum_{j=1}^{N}\frac{y_{j}}{A_{j}}\left[-p'_{i}\vec{v}_{j,j-1}\cdot\vec{v}_{j,j-1}-p'_{j}\vec{v}_{j-1,i}\cdot\vec{v}_{j,j-1}-p'_{j-1}\vec{v}_{i,j}\cdot \vec{v}_{j,j-1}\right]
\end{equation}

\noindent where rearranging the expression becomes
\begin{equation}\label{eq_grad_reor}
    (\nabla^{2}p')_{i}\sum_{j=1}^{N}y_{i}\frac{1}{3}=\frac{1}{4}\sum_{j=1}^{N}\frac{y_{j}}{A_{j}}\left[-(p'_{i}\vec{v}_{j,j-1}+p'_{j}\vec{v}_{j-1,i}+p'_{j-1}\vec{v}_{i,j})\cdot \vec{v}_{j,j-1}\right]
\end{equation}

\noindent Due to the vector \(\vec{v}_{j,j-1}=-(\vec{v}_{j-1,i}+\vec{v}_{i,j})\) , the previous expression can be rewritten as
\begin{equation}\label{eq_grad_reor_2}
    (\nabla^{2}p')_{i}\sum_{j=1}^{N}y_{i}\frac{1}{3}=\frac{1}{4}\sum_{j=1}^{N'}\frac{y_{j}}{A_{j}}\left\{\left[(p'_{i}-p'_{j})\vec{v}_{j-1,i}+[(p'_{i}-p'_{j-1})\vec{v}_{i,j}\right]\vec{v}_{j,j-1}\right\}
\end{equation}

\noindent Splitting the expression in parts corresponding to each edge
\begin{equation}\label{eq_grad_reor_2_1}
    (\nabla^{2}p')_{i}\sum_{j=1}^{N}y_{i}\frac{1}{3}=\frac{1}{4}\sum_{j=1}^{N}\frac{y_{j}}{A_{j}}\left\{(p'_{i}-p'_{j})\vec{v}_{j-1,i}\cdot\vec{v}_{j,j-1}\right\}+\frac{1}{4}\sum_{j=1}^{N}\frac{y_{j}}{A_{j}}\left\{(p'_{i}-p'_{j-1})\vec{v}_{i,j}\cdot\vec{v}_{j,j-1}\right\}
\end{equation}

\noindent It is assigned temporarily $j\rightarrow j'+1$ in order to group the difference in the variables of each part %% Figure!
\begin{equation}\label{eq_grad_reor_2_2}
    (\nabla^{2}p')_{i}\sum_{j=1}^{N}y_{i}\frac{1}{3}=\frac{1}{4}\sum_{j=1}^{N}\frac{y_{j}}{A_{j}}\left\{(p'_{i}-p'_{j})\vec{v}_{j-1,i}\cdot\vec{v}_{j,j-1}\right\}+\frac{1}{4}\sum_{j'=0}^{N-1}\frac{y_{j'+1}}{A_{j'+1}}\left\{(p'_{i}-p'_{j'})\vec{v}_{i,j'+1}\cdot\vec{v}_{j'+1,j'}\right\}
\end{equation}

\noindent As a result of the total of the summation giving the same result independently of the sum order of the terms, an offset that allows the correct definition of the contribution of each edge that connects the central node $i$ to the outward node is to be defined. Therefore, it can be considered for the new summation $j\rightarrow j'$.

\noindent Defining the area of the triangles as
\begin{subequations}\label{areas_t}
\begin{equation}
    A_{j}=\vec{v}_{j-1,i}\times \vec{v}_{j,j-1}=\frac{1}{2}|\vec{v}_{j-1,i}||\vec{v}_{j,j-1}|sin\beta_{j-1}
\end{equation}
\begin{equation}
    A_{j'+1}=\vec{v}_{i,j'+1}\times \vec{v}_{j'+1,j'}=\frac{1}{2}|\vec{v}_{i,j'+1}||\vec{v}_{j'+1,j'}|sin\alpha_{j'+1}
\end{equation}
\end{subequations}

\noindent and introducing the expression of the dot product 
\begin{subequations}\label{cdot_t}
\begin{equation}
    \vec{v}_{j-1,i}\cdot\vec{v}_{j,j-1}=|\vec{v}_{j-1,i}||\vec{v}_{j,j-1}|sin\beta_{j-1}
\end{equation}
\begin{equation}
    \vec{v}_{i,j'+1}\cdot\vec{v}_{j'+1,j'}=|\vec{v}_{i,j'+1}||\vec{v}_{j'+1,j'}|sin\alpha_{j'+1}
\end{equation}
\end{subequations}

\noindent Finally the Equation \ref{eq_grad_reor_2_2} becomes
\begin{equation}\label{eq_grad_reor_2_3}
    (\nabla^{2}p')_{i}\sum_{j=1}^{N}y_{i}\frac{1}{3}A_{j}=\sum_{j=1}^{N}(p'_{i}-p'_{j})\frac{1}{2}cot\beta_{j-1}+\sum_{j'=0}^{N-1}(p'_{i}-p'_{j'})\frac{1}{2}cot\alpha_{j'+1}
\end{equation}

\noindent Reassigning the indexes $j\rightarrow j'$ in the final expression of the Laplacian results in 
\begin{equation}\label{final_Laplacian}
    (\nabla^{2}p')_{i}=\frac{1}{\sum_{j=1}^{N}y_{i}\frac{A_{j}}{3}}\sum_{j=1}^{N}(p'_{i}-p'_{j})\left[\frac{1}{2}y_{j}cot\beta_{j-1}+\frac{1}{2}y_{j+1}cot\alpha_{j+1}\right]
\end{equation}

\noindent where \(y_{j}=(y_{i}+y_{j-1}+y_{j})/3\) and \(y_{j+1}=(y_{i}+y_{j}+y_{j+1})/3\), which are the radius of the centroids. Alternatively, the radius of the means of the middle points of each contour can be expressed as \(y_{j}=(5y_{i}+2y_{j-1}+5y_{j})/12\) and \(y_{j+1}=(5y_{i}+2y_{j}+5y_{j+1})/12\).

\section{Numerical implementation}

\subsection{System model}
Regarding the Equation \ref{k_entrekstar}, the software calculates the volumetric integral of the modal space of the combustion chamber and the surface integral of the modal space of the burning surface and nozzle entrance. Moreover, the complex values of the response functions are calculated based on the frequency of the acoustic mode. There is the option to quit the influence of the nozzle damping. Finally, the complex second-order equation is calculated and the stability value is obtained.

\subsection{Acoustic computation}
\noindent In order to compute the previous problem, the structure data is based on reference edges that connect the central node with the outward node. Connectivity matrices are used in order to set the nodes and triangles according to the reference edge. The following schemes establish the mesh topology. Therefore, it is necessary to ensure that the indexes are pointing at the correct element.

\begin{figure}[H]
    \includegraphics[height=5cm]{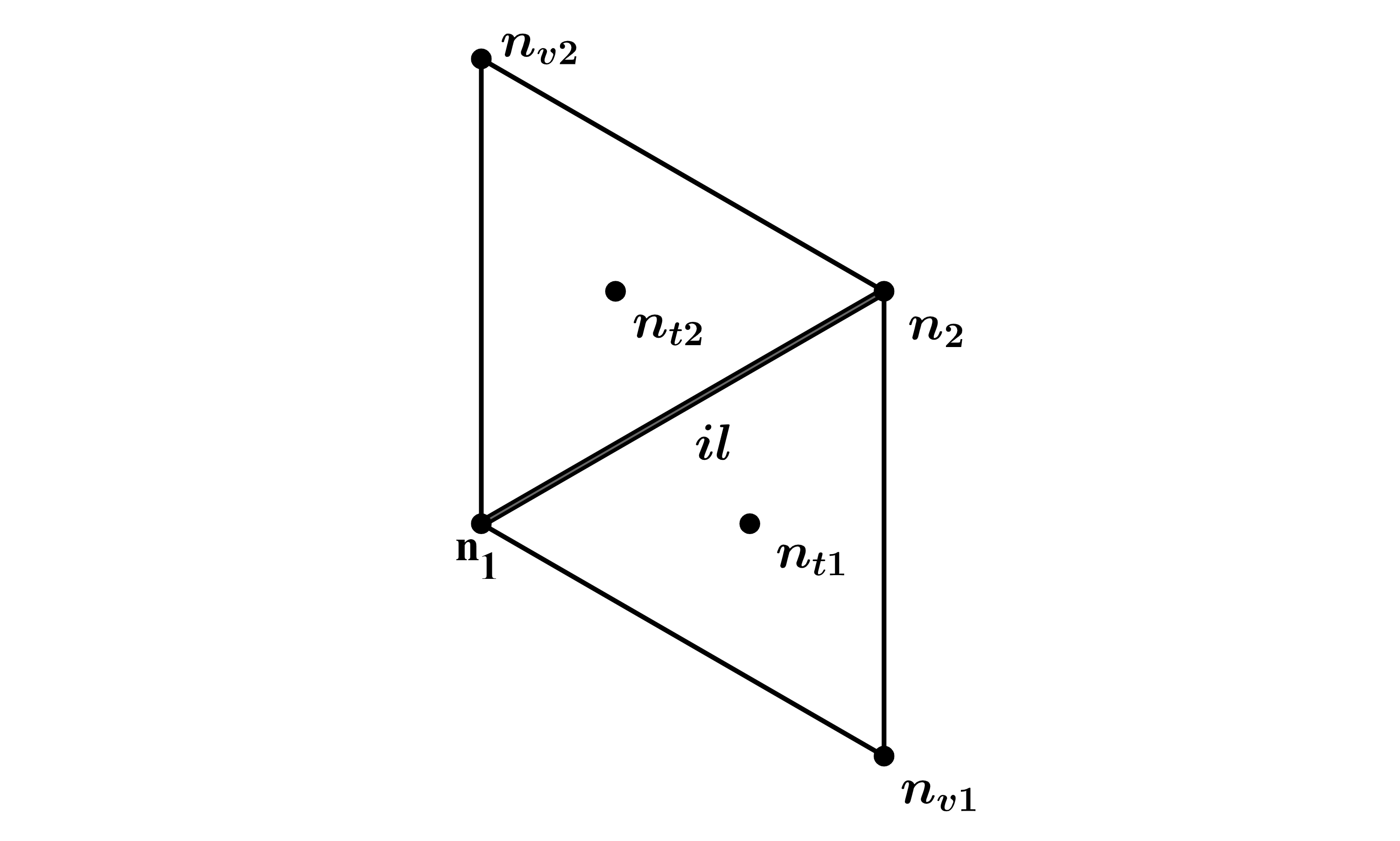}
    \centering
    \caption{Data structure.}
    \label{data_mesh}
\end{figure}

\noindent Once the correspondence between the different elements is established, it is possible to draw up the geometry.  The minimal control volume of the mesh is displayed in the Figure \ref{data_mesh}.\\

\noindent The program requires an input file created by \textit{MeshGen} \cite{tizón2023trimpackunstructuredtriangularmesh} that contains the information relative to the mesh of the cavity. It also contains a reference number for the contours of the section to set its character. The user has to establish if the problem is bidimensional or axisymmetric. In case the problem is axisymmetric, the program requires the specification of the Bessel wave numbers $m$.

\subsubsection{Connectivity matrix}

\noindent The connectivity matrices are performed by a given triangulation using the function \textit{connec} that can be found in the library \cite{tizón2023trimpackunstructuredtriangularmesh} \textit{TRIMPACK} programmed in FORTRAN. However, its algorithm was programmed in C++ in order to avoid problem compatibilities between C++ and FORTRAN. Three types of connectivity matrices are used: nodes as a function of triangles, triangles as a function of edges, and nodes as a function of edges.  \\

\noindent Another connectivity matrix algorithm is programmed in order to associate each edge with the third node of each of the triangles that share the edge as a side. Moreover, the matrix contains the information if one of the nodes doesn't exist. Therefore, it indicates that the edge belongs to the contour of the mesh. 

\subsubsection{Sparse matrix}

\noindent The sparse matrix contains the values of the Laplacian discretization in order to obtain its eigenvalues and eigenvectors. The first step is to obtain a vector that contains the number of triangles connected to each node. In case the problem is axisymmetric, another vector is computed to set which nodes belong to the axis. \\

\noindent Another matrix is computed in order to calculate the values of the \textit{cotangent factors} depending on the geometry and the radius model chosen by the user. The area of the triangles is calculated by solving the determinant of the plane. Therefore, the area corresponding to the control volume  associated with each central node is computed regarding if the case is axisymmetric.\\

\noindent Finally, the values of the sparse matrix are generated. In case the problem is axisymmetric, the factor $m^2/r^2$ is added to the diagonal of the matrix. Furthermore, a loop over the edges is done, computing the contribution of the different outward nodes to the central node depending on the radius model that has been chosen. \\

\noindent In the axisymmetric case, the radius of the nodes at the axis is equal to zero. Nevertheless, the software computes the radius as a means of delimiting the centroids of the triangles by the two middle sides, the contour, and the segment which connects the centroid of the principal triangle and the central node. Due to the area of each minimal cell being calculated by dividing the area of the triangle by three, this contribution cannot be eliminated. In the case of the node at the axis, the contribution of its area to the Laplacian operator is infinite. This contradicts the boundary condition established in the theory framework defining the variations of fluctuations of pressure at the axis as a real number. Therefore, to subtract the contribution that tends to infinity, the contribution of the outward nodes that belong to the axis to the central node at the axis is eliminated from the sparse matrix, and only the contribution of the outward nodes that do not belong to the axis is to be considered.\\

\subsubsection{Eigenvalues and eigenvectors}

Once the sparse matrix has been obtained, the program calculates its eigenvalues and eigenvectors via the Spectra Library (Sparse Eigenvalue Computation Toolkit as a Redesigned ARPACK) \cite{Spectra}. It is a C++ library for large-scale eigenvalue problems, built on top of \textit{Eigen}, an open-source linear algebra library.\\

\noindent In this case, it is a large matrix where only a few coefficients are different from zero. Therefore, the memory consumption can be reduced, increasing the performance by using a specialized representation storing only the nonzero coefficients. The data provided by the program are three vectors which indicate the position and the value at the sparse matrix.

\subsubsection{Eigen Solvers}
Once the sparse matrix has been defined, a specified number \textit{k} of its eigenvalues, indicated by the user, is calculated. In this study case, only the eigenvalues with the lowest magnitude are interesting to compute. 
The program has two options for Eigen Solvers. For the majority of cases, its time computing is relatively similar despite their different algorithms. The \textit{GenEigsSolver} computes the problem \(Ax=\lambda x\) for general real matrices. \\

\noindent On the other hand, the \textit{GenEigsRealShiftSolver} computes the problem for general real matrices with a real shift value in the shift-and-invert mode. This mode is based on the following fact: If $\lambda$ and $x$ are a pair of eigenvalue and eigenvector of matrix $A$, such that $Ax=\lambda x$, then for any $\sigma$, \((A-\sigma I)^{-1}x=\nu x\) where \(\nu = 1/(\lambda - \sigma)\) which indicates that ($\nu$,x) is an eigenpair of the matrix \((A-\sigma I)^{-1}\). Therefore, if it is passed the matrix operation \((A-\sigma I)^{-1}y\) (rather than $Ay$) to the eigen solver, then it would get the desired values of $\nu$, and $\lambda$ can also be easily obtained by noting that $\lambda=\sigma+\nu^{-1}$. \\

\noindent The reason why it needs this type of manipulation is that the algorithm of \textit{SPECTRA} (and also \textit{ARPACK}) is good at finding eigenvalues with large magnitude, but may fail in looking for eigenvalues that are close to zero. However, it can be set \(\sigma =0 \), find the largest eigenvalues of $A^{-1}$, and then transform back to $\lambda$, since in this case largest values of $\nu$ imply the smallest values of $\lambda$.

\section{Results}

\subsection{Acoustic computation}

The results given by the program have been compared with the analytical solutions. The analytical eigenmodes have been plotted in an isometric and contour view. To display the eigenmode solutions given by the program, a high-level plotting library for displaying data called DISLIN \cite{DISLIN} has been used.\\

\subsubsection{Validation case: Axisymmetric case}

\begin{figure}[H]
    \includegraphics[height = 2.7cm, width = 0.65\textwidth]{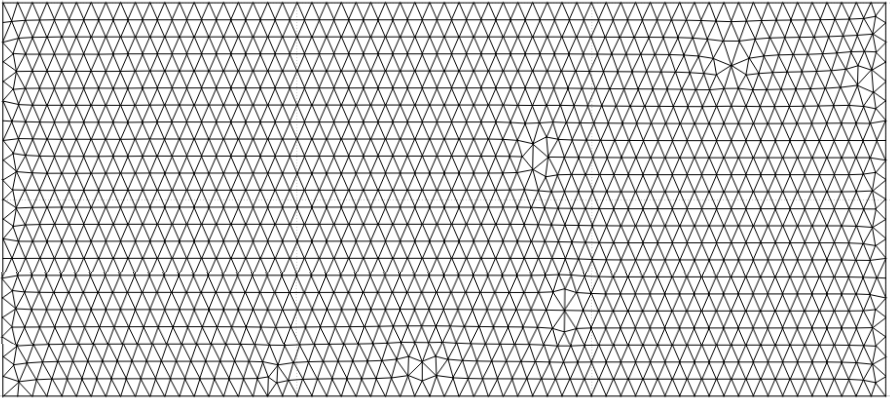}
    \centering
    \caption{Representation of the bi-dimensional geometry mesh data}
    \label{mesh2d}
\end{figure}

The geometry mesh data is displayed in Figure \ref{mesh2d}. In this case, the Laplacian operator is computed as axisymmetrical, and, due to the interest of our study, the solution has been obtained for Newman boundary conditions. The first six eigenvalues and their residuals can be seen in Table \ref{eigenmodes_cil}. The residuals have been calculated as the norm of the columns of the matrix \([A\cdot v - \lambda \cdot v]\), where $v$ and $\lambda$ are the eigenvectors and eigenvalues respectively, divided by the values of the corresponding eigenvalue. \\

\begin{table}[H]
  \centering
    \begin{tabular}{c | c c }
    \hline
    (n,m,l) & Value & Residual\\
    \hline
    (0,0,1) & 1.09643 & 3.14388e-07 \\
    (0,1,0) & 3.38869 & 1.38071e-13 \\
    (0,0,2) & 4.38346 & 2.6806e-07 \\
    (0,1,1) & 4.48475 & 1.06629e-13 \\
    (0,1,2) & 7.77058 & 6.88415e-14 \\
    (0,2,0) & 40.3165 & 1.85053e-09 \\
    \hline
    \end{tabular}%
    \caption{Values and residuals of the first six pressure eigenmodes in the transverse plane of a cylinder with rigid walls.}
  \label{eigenmodes_cil}%
\end{table}%

\begin{figure}[H]
    \includegraphics[height = 3.5cm, width = 0.85\textwidth]{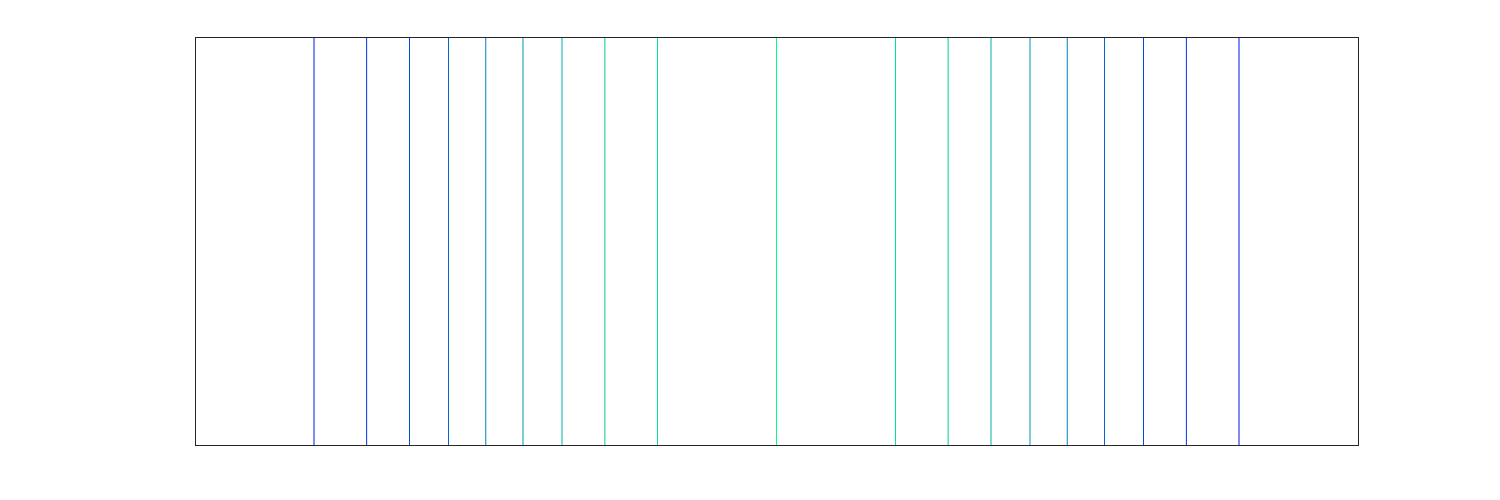}
    \centering
    \caption{Analytical contour plane representation of the pressure eigenmode (2,0,0) in the transverse plane of a cylinder with rigid walls in cartesian coordinates.}
    \label{plane_repre_contour}
\end{figure}

\begin{figure}[H]
    \includegraphics[width = 0.7\textwidth]{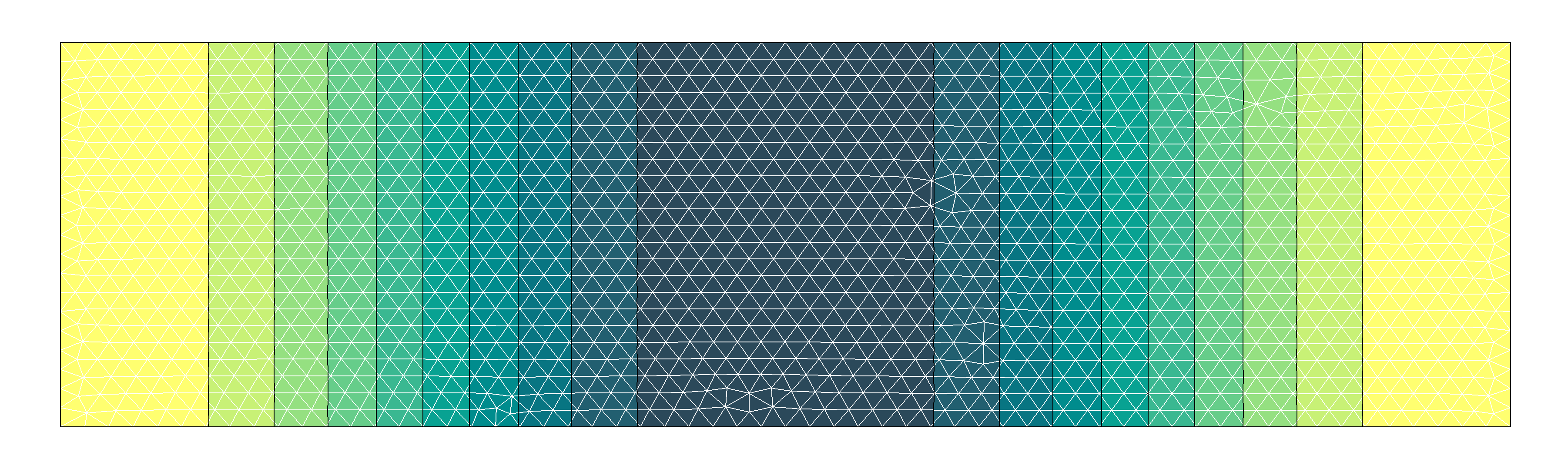}
    \centering
    \caption{Numerical contour representation of the pressure eigenmode (2,0,0) in the transverse plane of a cylinder with rigid walls and $1/\sqrt{N}=0.026$ in cartesian coordinates. }
    \label{contour_mode3}
\end{figure}

\noindent Moreover, Figure \ref{plane_repre_contour} and Figure \ref{contour_mode3} compare the analytical and numerical representation of one of the modes, showcasing a good similarity. Usually, the eigenmodes of a cylindrical cavity are represented in a circular plane. Nevertheless, due to the formulation used to discretise the Laplacian, the eigenmodes of the solution have to be represented in rectangular mesh data, corresponding to a longitudinal semi-section of the cylinder. Therefore, the comparison between the contour plots has been done with the analytical solution plotted with the previous reasoning. \\

\begin{figure}[H]
    \includegraphics[width = 0.7\textwidth]{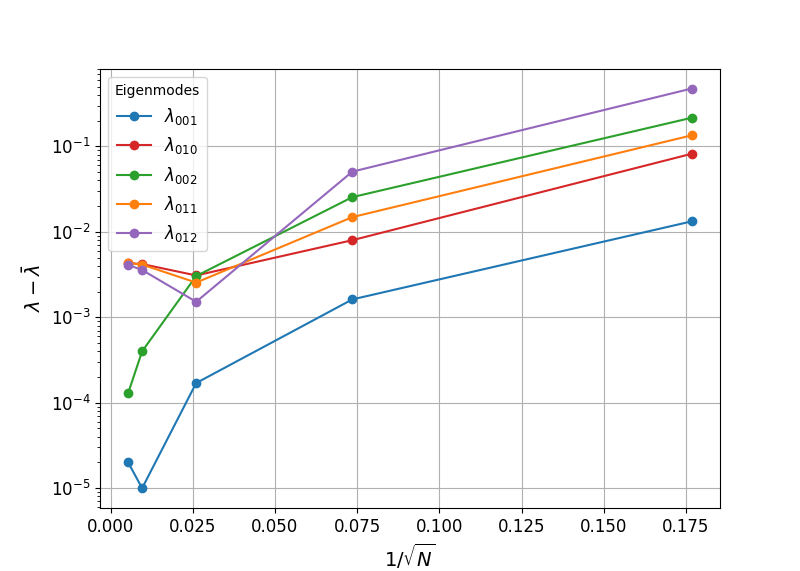}
    \centering
    \caption{Eigenvalues mesh convergence.}
    \label{eigenval_cil}
\end{figure}

\begin{figure}[H]
    \includegraphics[width = 0.7\textwidth]{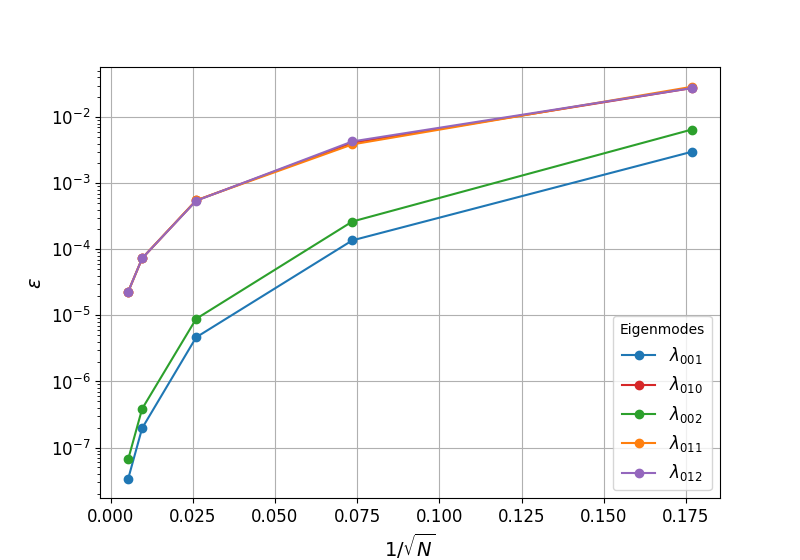}
    \centering
    \caption{Laplacian operator mesh convergence }
    \label{eigenvec_cil}
\end{figure}

\noindent Finally, a mesh convergence study of the Laplacian operator has been carried out. The error of the Laplacian operator has been defined as

\begin{equation}\label{eigenvecerror}
    \varepsilon = \frac{1}{N}\sqrt{\sum_{i=1}^{N}(\nabla^2f-\nabla^2\Bar{f})}
\end{equation}

\noindent where \(\nabla^2f\) represents the numerical solution and \(\nabla^2\Bar{f}\) the analytical solution. It is to be noted that the eigenvectors calculated by \textit{SPECTRA} library have been normalized with the Euclidean norm equal to unity. Therefore, in order to obtain the previously defined error, the analytical solution has been normalized too.  In Figure \ref{eigenval_cil} and Figure \ref{eigenvec_cil} a mesh convergence study is presented. A good convergence for a mesh of \(1/\sqrt{N} \approx 0.03\) is observed. Moreover, it can be concluded from the Laplacian operator mesh convergence, that the eigenmodes corresponding to $m=0$ present an error \(10^-2\) lower than the rest of eigenmodes with a higher $m$ and with \(1/\sqrt{N} \approx 0.03\). 

\subsubsection{Solid rocket motor geometry}

\begin{figure}[H]
    \includegraphics[height = 3cm, width=0.7\textwidth]{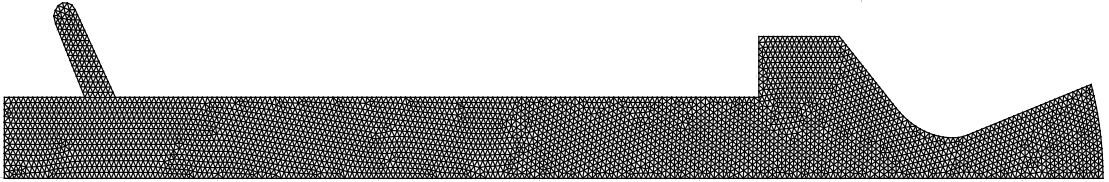}
    \centering
    \caption{Representation of the rocket geometry mesh data.}
    \label{meshrocket}
\end{figure}

The pressure eigenmodes of the cavity more similar to a solid rocket engine have been calculated. Figure \ref{meshrocket}  shows the axis segment, the reacting walls, and the nozzle. In this study case, the nozzle is modelled only with the convergent entrance of the nozzle. Moreover, Table \ref{eigenmodes_meshrocket} displays the first six modes and their residuals of the geometry.

\begin{table}[H]
  \centering
    \begin{tabular}{c | c c }
    \hline
    (n,m) & Value & Residual\\
    \hline
    (1,0) & 17.1065 & 4.07962e-05 \\
    (2,0) & 67.1392 & 6.45383e-05 \\
    (3,0) & 109.109 & 1.40561e-04 \\
    (4,0) & 263.042 & 5.22578e-05 \\
    (5,0) & 515.265 & 6.78284e-05 \\
    (1,1) & 661.530 & 1.95696e-13 \\
    \hline
    \end{tabular}%
    \caption{Values and residuals of the first six pressure eigenmodes of a mesh rocket geometry.}
  \label{eigenmodes_meshrocket}%
\end{table}%

\begin{figure}[H]
    \includegraphics[width = 0.7\textwidth]{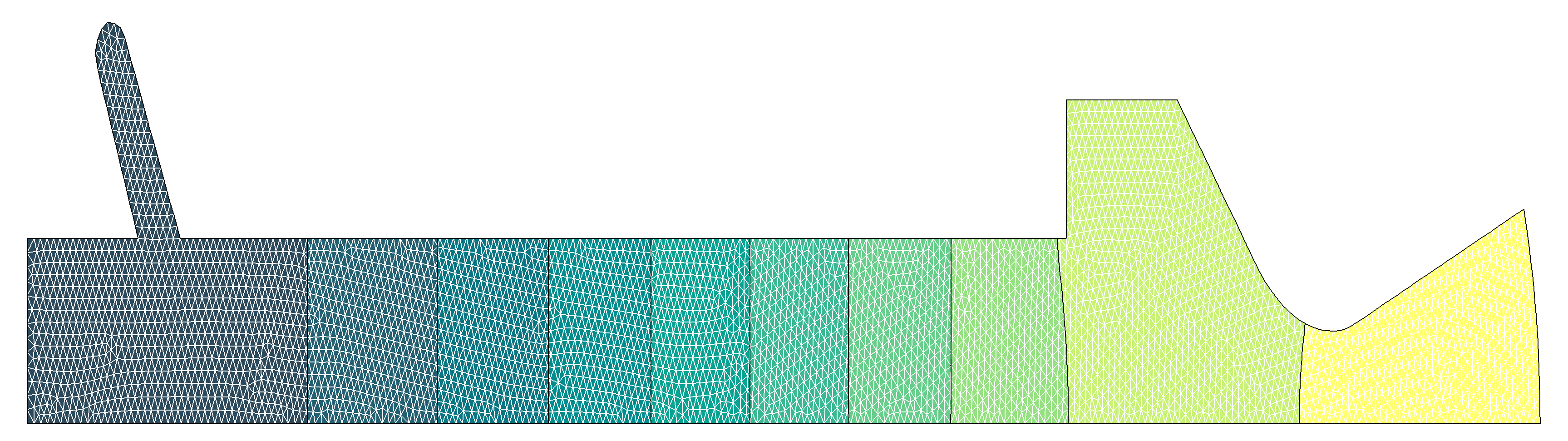}
    \centering
    \caption{Numerical contour representation of the $1^{st}$ pressure eigenmode in the transverse plane of a cylinder with rigid walls.}
    \label{rock_1}
\end{figure}
\begin{figure}[H]
    \includegraphics[width = 0.7\textwidth]{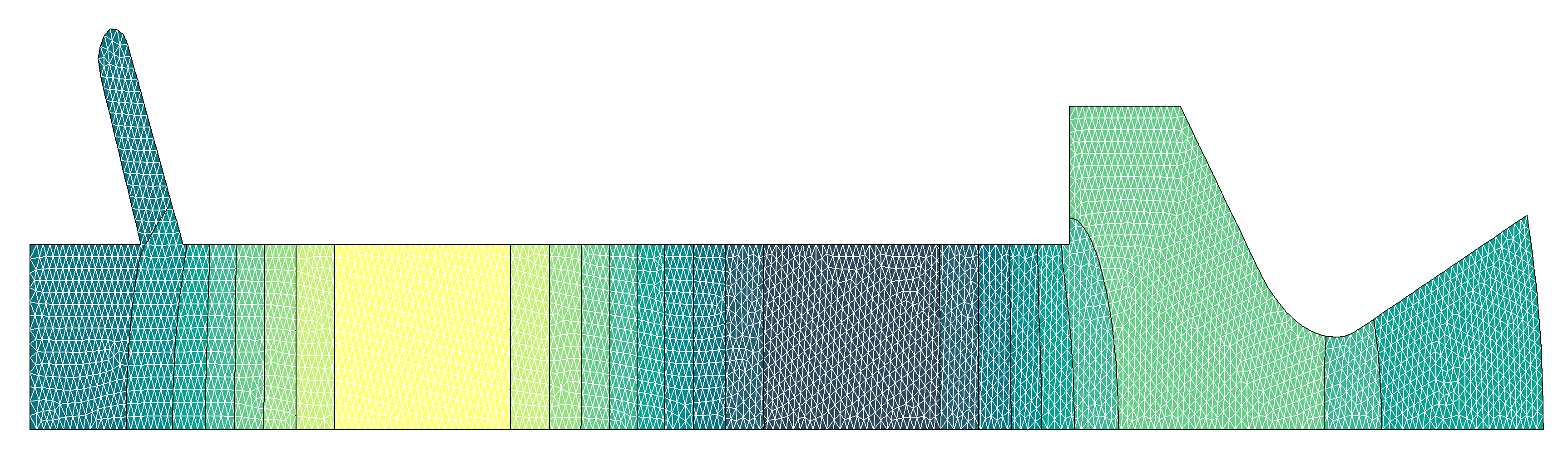}
    \centering
    \caption{Numerical contour representation of the $4^{th}$ pressure eigenmode in the transverse plane of a cylinder with rigid walls.}
    \label{rock_4}
\end{figure}
\begin{figure}[H]
    \includegraphics[width = 0.7\textwidth]{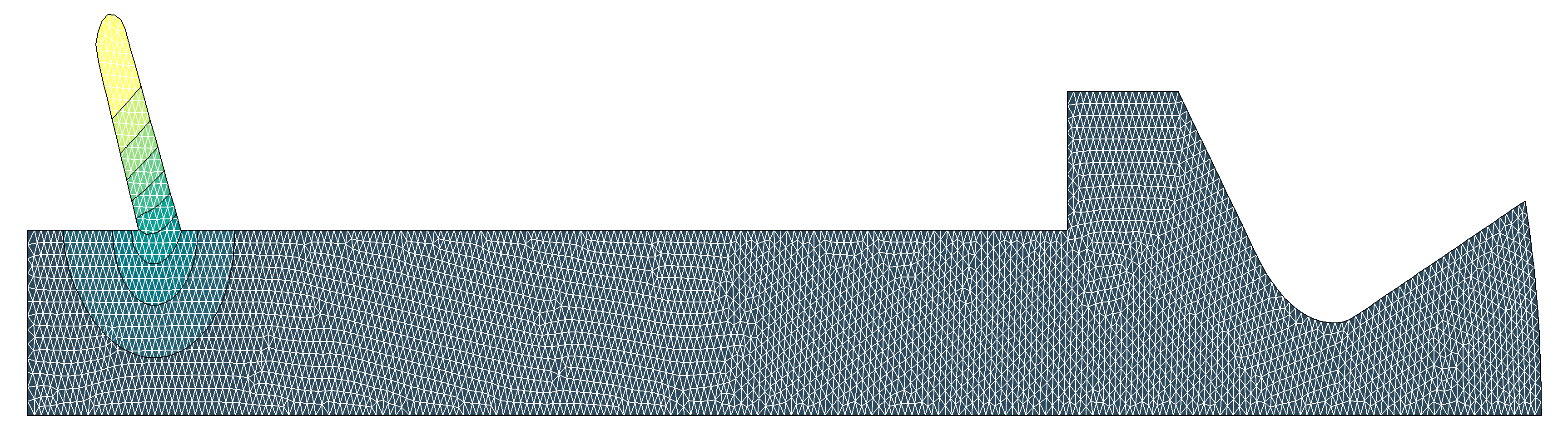}
    \centering
    \caption{Numerical contour representation of the $6^{th}$ pressure eigenmode in the transverse plane of a cylinder with rigid walls.}
    \label{rock_6}
\end{figure}

\noindent In Figures \ref{rock_1}-\ref{rock_6} different eigenmode representations are displayed. In Figure \ref{rock_1} can be observed that the first eigenmode corresponds to longitudinal mode. Meanwhile, the sixth mode corresponds to a transverse mode in the slot region of the geometry.

\subsection{Stability analysis}

\noindent In this section different stability results obtained from the software developed will be commented on, based on the theoretical framework described previously, while changing different variables of the model. To compare the results, different rocket motor parameters have been set, which cover the stability problem

\begin{table}[H]
  \centering
    \begin{tabular}{r | c }
    \hline
    Parameter & Value\\
    \hline
    $\dot{r}_b$ [m/s] & 0.01 \\
    \hline
    $\alpha_p$ [$m^2$/s] & $10^{-7}$ \\
    \hline
    $c^*$ [m/s] & 1100 \\
    \hline
    $\gamma$ & 1.2 \\
    \hline
    n & 0.3 \\
    \hline
    $n_s$ & 0 \\
    \hline
    $M_1$ & 0.3\\
    \hline
    $D_g$ [m] & 0.02 \\
    \hline
    \end{tabular}%
    \caption{Propellant and motor parameters.}
  \label{eigenmodes_fixedrim}%
\end{table}%

\noindent where $\dot{r}_b$ is the regression rate, $\alpha_p$ is the diffusivity of the propellant, $\gamma$ is the relation of specific heats, $n$ is the Vieille parameter, $n_s$ is a pyrolysis factor, $M_1$ is the Mach number at the entrance of the nozzle and $D_g$ is the diameter of the nozzle throat.

\subsubsection{Effect of the Response Function}

\begin{figure}[H]
    \hspace{1cm}
    \begin{subfigure}{7cm}
        \includegraphics[height = 0.8cm, width=0.8\textwidth]{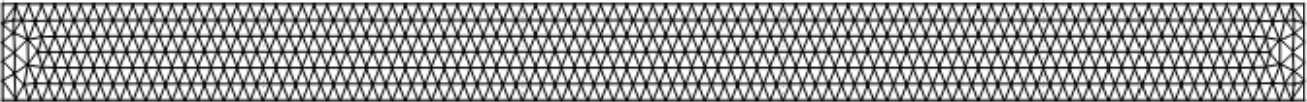}
        \centering
        \caption{Charge with $D_c$=0.05m.}
        \label{cil_charge}
    \end{subfigure}
    \hfill
    \begin{subfigure}{7cm}
        \includegraphics[height = 1.6cm, width=0.8\textwidth]{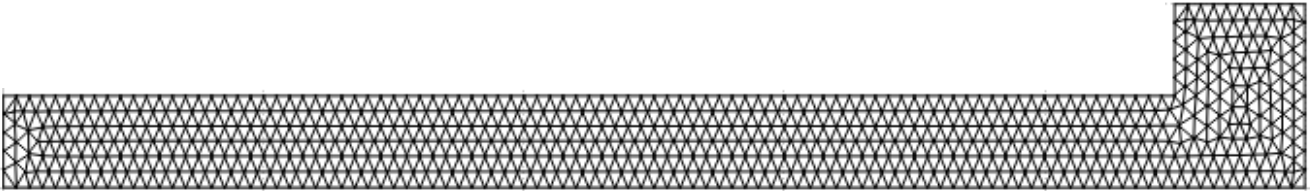}
        \centering
        \caption{Charge with combustion of the bases and $D_c$=0.1m.}
        \label{cil_charge_dc}
    \end{subfigure}
    \hspace{1cm}
    \begin{center}
    \begin{subfigure}{15cm}
        \includegraphics[height = 2.5cm, width=0.4\textwidth]{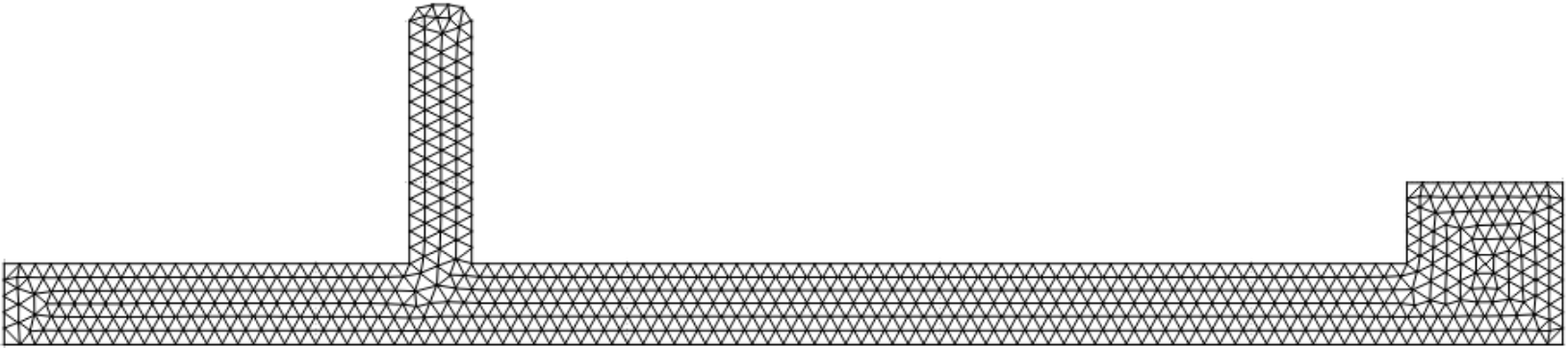}
        \centering
        \caption{Slotted charge with the combustion of the bases and $D_c$=0.1m.}
        \label{cil_charge_dc_slot}
        \end{subfigure}
    \end{center}
    \caption{Different cylindrical geometries of combustion chambers.}
    \label{geo_chambers}
\end{figure}

\noindent For this study, three different geometries will be used. The first one is a cylindrical charge, the second geometry has combustion in the bases and the diameter of the convergent entrance ($D_c$) has been increased. Finally, the third one presents a slot in order to increase the burning surface. The geometries of the Figure \ref{geo_chambers} are semi-sections of 2D axisymmetric combustion chambers. \\

\noindent Having calculated the first ten modes of each cavity, its stability value is displayed versus the dimensionless frequency. Moreover,  the response function of the burning surface has been plotted, in order to better visualize its influence on the stability of the modes. In order to observe the influence of the nozzle response function, the stability value has been calculated without the presence of the nozzle response function too. The parameters of the burning surface response function have been set to $A=7$ and $B=0.6$, and the limit value of the dimensionless frequency of the nozzle response function is $\Omega_c = 100$. 

\begin{figure}[H]
    \includegraphics[width = 0.7\textwidth]{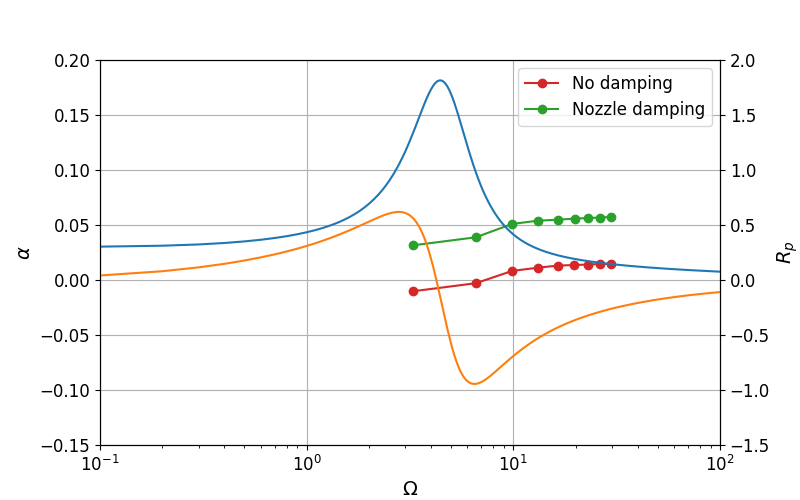}
    \centering
    \caption{Representation of the stability value of the first tenth modes of the combustion chamber \ref{cil_charge}.}
    \label{cil_charge_1_d}
\end{figure}

\noindent The Table \ref{tab:cil_t1_z} displays, for each acoustic mode, the frequency, the Bessel number, the value of the dimensionless integrals of the burning surface and nozzle entrance, and the stability value.

\begin{table}[H]
  \centering
    \begin{tabular}{c |c |c |c |c |c}
    \hline
    mode & $\omega$ & $m$ & {\footnotesize $\Tilde{J}_p$} & {\footnotesize $\Tilde{J}_g$} & $\alpha$\\
    \hline
    1     & 3292.61 & 0     & 39.9971 & 2.00034 & 0.0355986 \\
    2     & 6584.63 & 0     & 39.9884 & 2.00136 & 0.0221155 \\
    3     & 9875.47 & 0     & 39.974 & 2.00306 & 0.0476327 \\
    4     & 13164.5 & 0     & 39.9537 & 2.00545 & 0.0523414 \\
    5     & 16451.2 & 0     & 39.9276 & 2.00854 & 0.0540422 \\
    6     & 19735 & 0     & 39.8957 & 2.01235 & 0.0549415 \\
    7     & 23015.2 & 0     & 39.8579 & 2.01689 & 0.0555333 \\
    8     & 26291.3 & 0     & 39.8143 & 2.02218 & 0.055981 \\
    9     & 29562.6 & 0     & 39.7647 & 2.02827 & 0.056353 \\
    \hline
    \end{tabular}%
    \caption{Values of the stability study of the combustion chamber \ref{cil_charge} with nozzle damping.}
  \label{tab:cil_t1_z}%
\end{table}%

\noindent Considering only the stability values based on the acoustic modes, it can be appreciated in Figure \ref{cil_charge_1_d} how, depending on the damping, the first and second modes present a negative instability value. Therefore, the combustion chamber will be susceptible to suffering combustion instabilities. On the other hand,  a constant evolution of the stability values for each mode is observed, as they present the same type of mode, the longitudinal one. Therefore, the integrals over the spatial mode present similar values.  It can also be observed that the nozzle transfer function has a similar effect on each mode, due to the value of the integral over the nozzle entrance not differing significantly between modes.

\begin{figure}[H]
    \includegraphics[width = 0.7\textwidth]{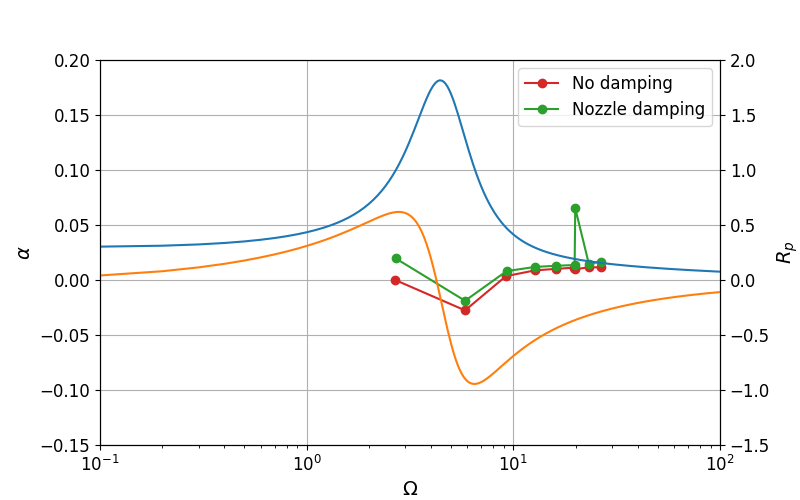}
    \centering
    \caption{Representation of the stability parameter of the first tenth modes of the combustion chamber \ref{cil_charge_dc}.}
    \label{cil_charge_dc_1_c}
\end{figure}

\begin{table}[H]
  \centering
    \begin{tabular}{c |c |c |c |c |c}
    \hline
    mode & $\omega$ & $m$ & {\footnotesize $\Tilde{J}_p$} & {\footnotesize $\Tilde{J}_g$} & $\alpha$\\
    \hline
    1     & 2692.98 & 0     & 37.2233 & 3.76557 & 0.0198272 \\
    2     & 5851.16 & 0     & 40.2347 & 1.62493 & -0.0189632 \\
    3     & 9239.69 & 0     & 41.2606 & 0.873018 & 0.00801334 \\
    4     & 12705.6 & 0     & 41.5867 & 0.601731 & 0.0119564 \\
    5     & 16196 & 0     & 41.6415 & 0.509471 & 0.0131142 \\
    6     & 19687 & 0     & 41.5415 & 0.518831 & 0.0139038 \\
    7     & 19902.7 & 1     & 37.6382 & 10.652 & 0.0658184 \\
    8     & 23160.3 & 0     & 41.3007 & 0.639312 & 0.0149203 \\
    9     & 26589.4 & 0     & 40.8672 & 0.992157 & 0.0169461 \\
    \hline
    \end{tabular}%
    \caption{Values of the stability study of the combustion chamber \ref{cil_charge_dc} with nozzle damping.}
  \label{tab:cil_charge_dc_t1_q}%
\end{table}%

\begin{figure}[H]
    \hspace{1cm}
    \begin{subfigure}{7cm}
        \includegraphics[height = 2cm, width=\textwidth]{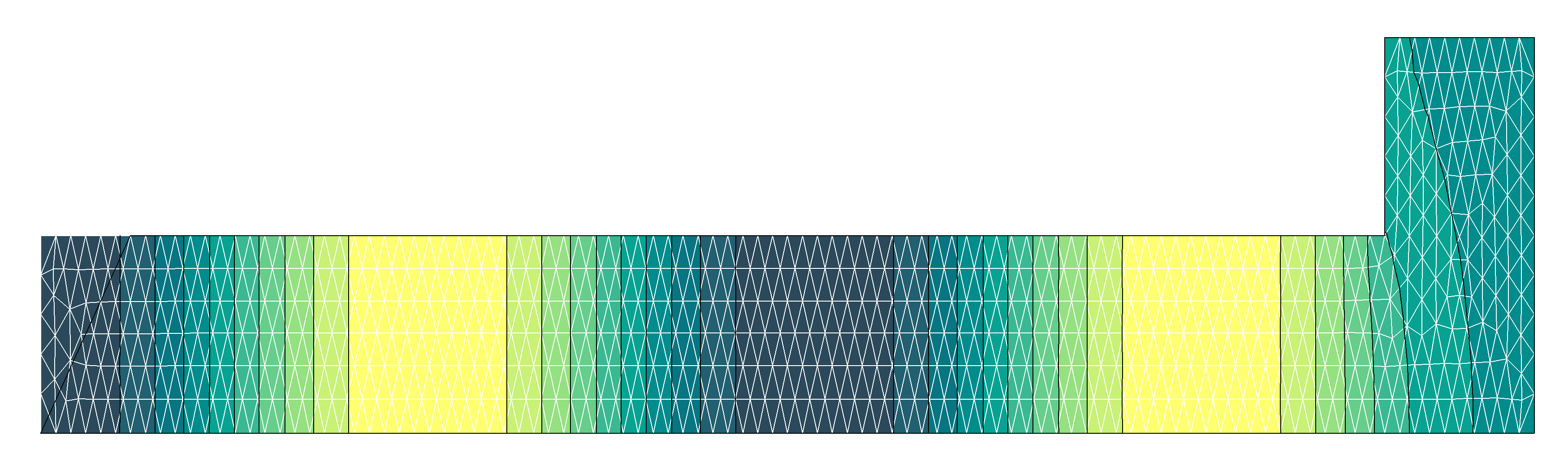}
        \centering
        \caption{$4^{th}$ mode.}
        \label{4mode_g}
    \end{subfigure}
    \hfill
    \begin{subfigure}{7cm}
        \includegraphics[height = 2cm, width=\textwidth]{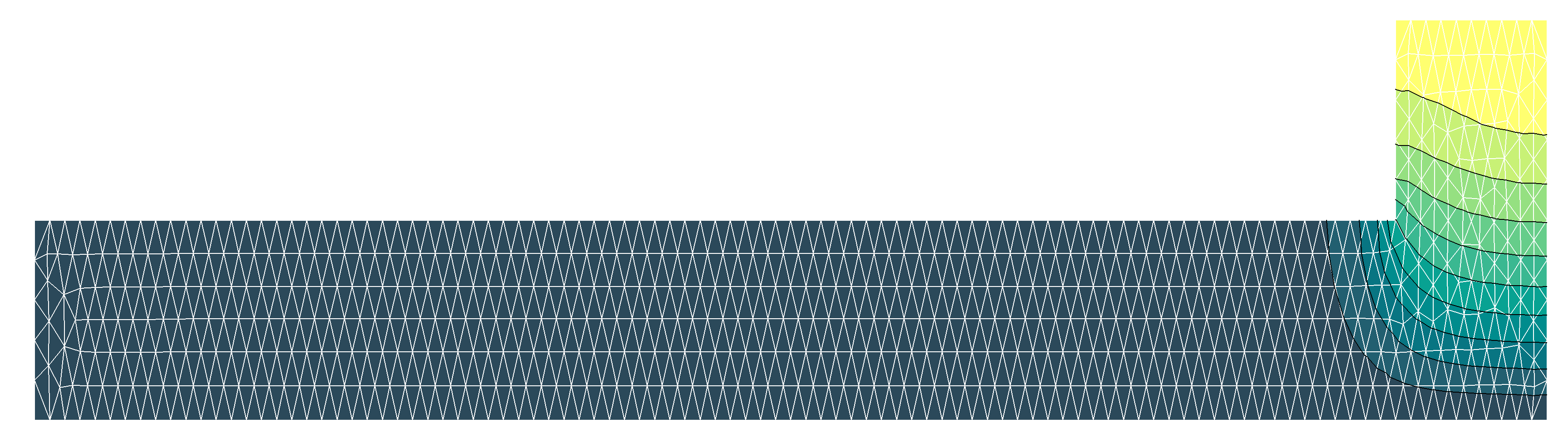}
        \centering
        \caption{$7^{th}$ mode.}
        \label{7mode_2_l}
    \end{subfigure}
    \hspace{1cm}
    \caption{Contour representation of the eigenmodes of the cavity \ref{cil_charge_dc}}
    \label{response_burning_2_t}
\end{figure}

\noindent In the case of Figure \ref{cil_charge_dc_1_c} it is observed that the evolution of the stability value between the modes is not as constant as in the previous combustion chamber. For the first mode,  the influence of the nozzle response function can be recognized because it is a longitudinal mode. For the following modes, it is clearly visible how low the influence of the nozzle transfer function is, as seen in Table \ref{tab:cil_charge_dc_t1_q}. The previous statement can be explained by analyzing how in Figure \ref{4mode_g}, a longitudinal and transverse mode confluence at the entrance nozzle region. Therefore, it is coherent with the statement that the nozzle absorbs more acoustic energy when the mode shapes are longitudinal.

\noindent The seventh mode shows a high influence of the nozzle damping. The stability value increases a lot due to the modal space being highly present near the nozzle entrance region. The reason is that the transverse mode is present just in the entrance nozzle region, as can be seen in Figure \ref{7mode_2_l}. Nevertheless, the previous result can not be seen as a conclusion, as it requires a better study in the theoretical framework of the influence of the nozzle damping depending on the mode shape type. The eighth and ninth modes present similar evolutions of the stability value to the previous ones. Therefore, it is observed that the damping of the nozzle does not significantly affect the majority of the modes of this combustion chamber.

\begin{figure}[H]
    \includegraphics[width = 0.6\textwidth]{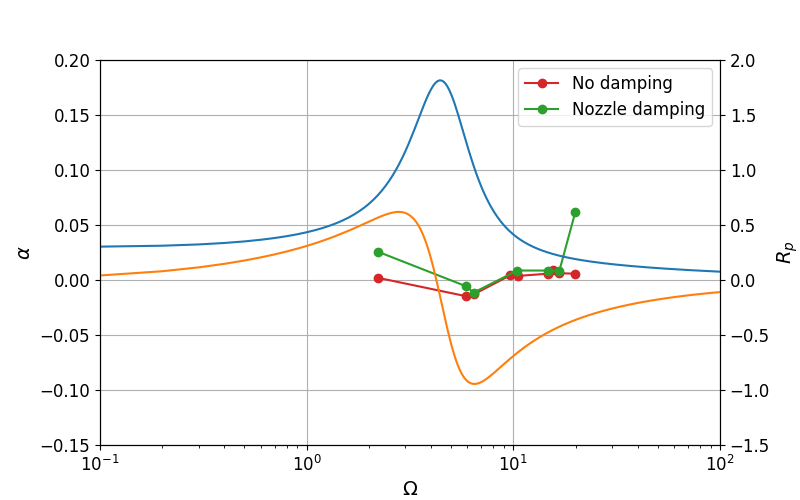}
    \centering
    \caption{Representation of the stability parameter of the first tenth modes of the combustion chamber \ref{cil_charge_dc_slot}.}
    \label{cil_charge_dc_slot_1_l}
\end{figure}

\begin{table}[H]
  \centering
    \begin{tabular}{c |c |c |c |c |c}
    \hline
    mode & $\omega$ & $m$ & {\footnotesize $\Tilde{J}_p$} & {\footnotesize $\Tilde{J}_g$} & $\alpha$\\
    \hline
    1     & 2198.62 & 0     & 39.0182 & 4.59041 & 0.0258097 \\
    2     & 5914.82 & 0     & 39.6878 & 1.76073 & -0.00571675 \\
    3     & 6411.75 & 0     & 50.3595 & 0.298616 & -0.011877 \\
    4     & 9642.65 & 1     & 61.6582 & 1.84E-21 & 0.00389385 \\
    5     & 10497.6 & 0     & 40.226 & 0.974577 & 0.00860065 \\
    6     & 14680.4 & 0     & 41.7924 & 0.496684 & 0.00825962 \\
    7     & 15436.6 & 2     & 64.0456 & 1.81E-27 & 0.008921 \\
    8     & 16680.9 & 0     & 43.5187 & 0.305363 & 0.00789059 \\
    9     & 19902.8 & 1     & 37.638 & 10.652 & 0.0614759 \\
    \hline
    \end{tabular}%
    \caption{Values of the stability study of the combustion chamber \ref{cil_charge_dc_slot} with nozzle damping.}
  \label{tab:cil_charge_dc_slot_t1_p}%
\end{table}%

\begin{figure}[H]
    \hspace{1cm}
    \begin{subfigure}{7cm}
        \includegraphics[height = 3cm, width = \textwidth]{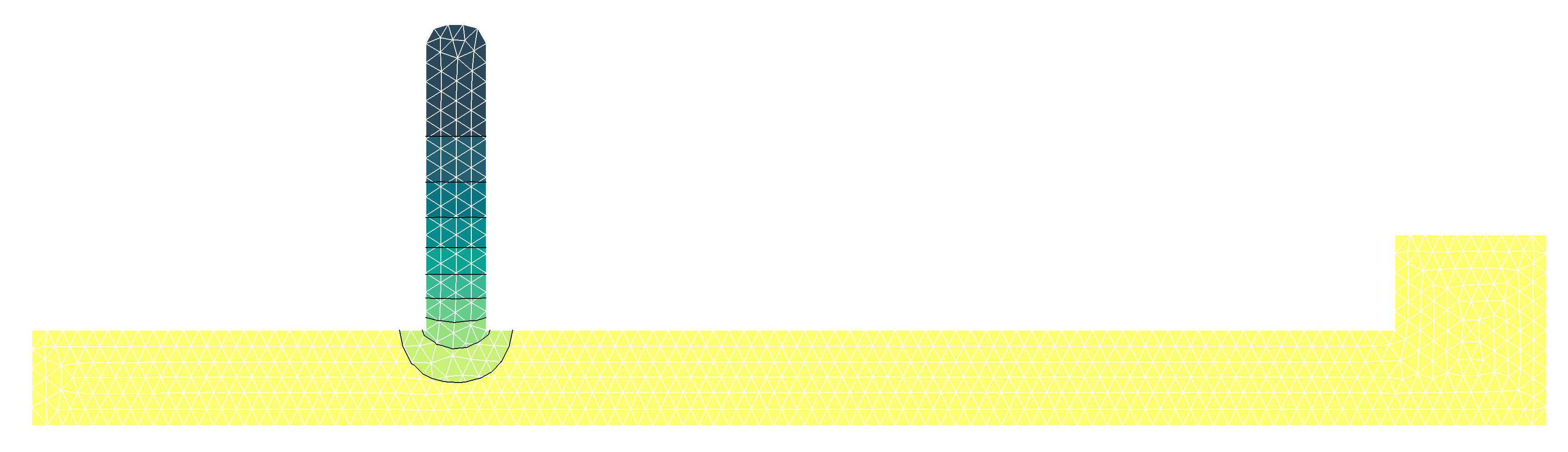}
        \centering
        \caption{$4^{th}$ mode.}
        \label{4mode_6}
    \end{subfigure}
    \hfill
    \begin{subfigure}{7cm}
        \includegraphics[height = 3cm, width = \textwidth]{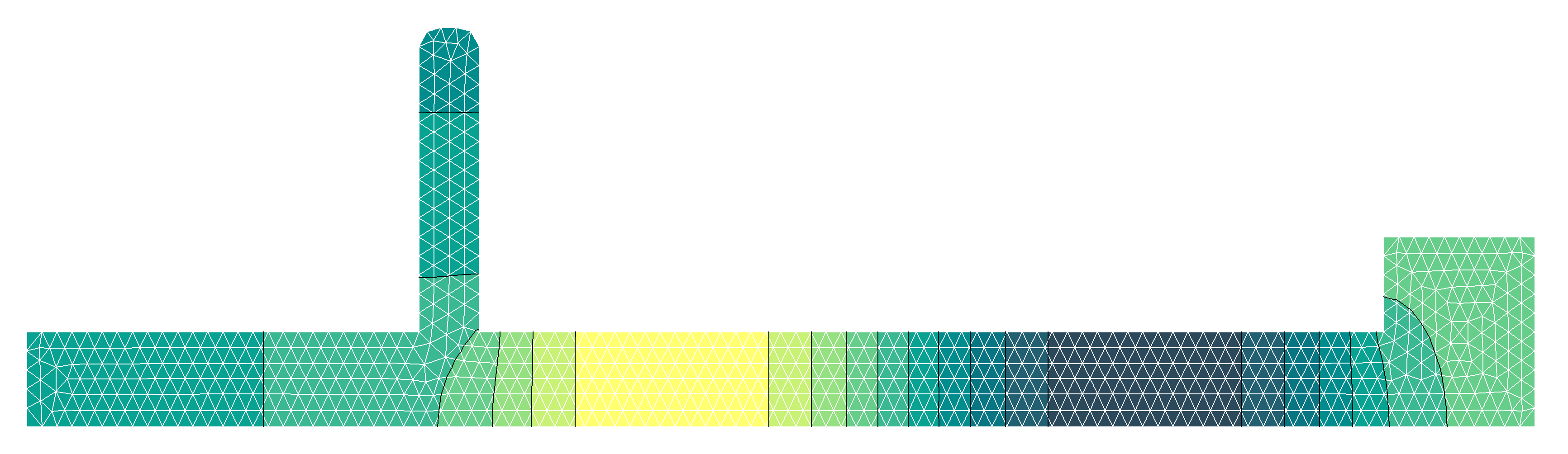}
        \centering
        \caption{$5^{th}$ mode.}
        \label{7mode_5}
    \end{subfigure}
    \hspace{1cm}
    \begin{center}
    \begin{subfigure}{8cm}
        \includegraphics[height = 3cm, width = \textwidth]{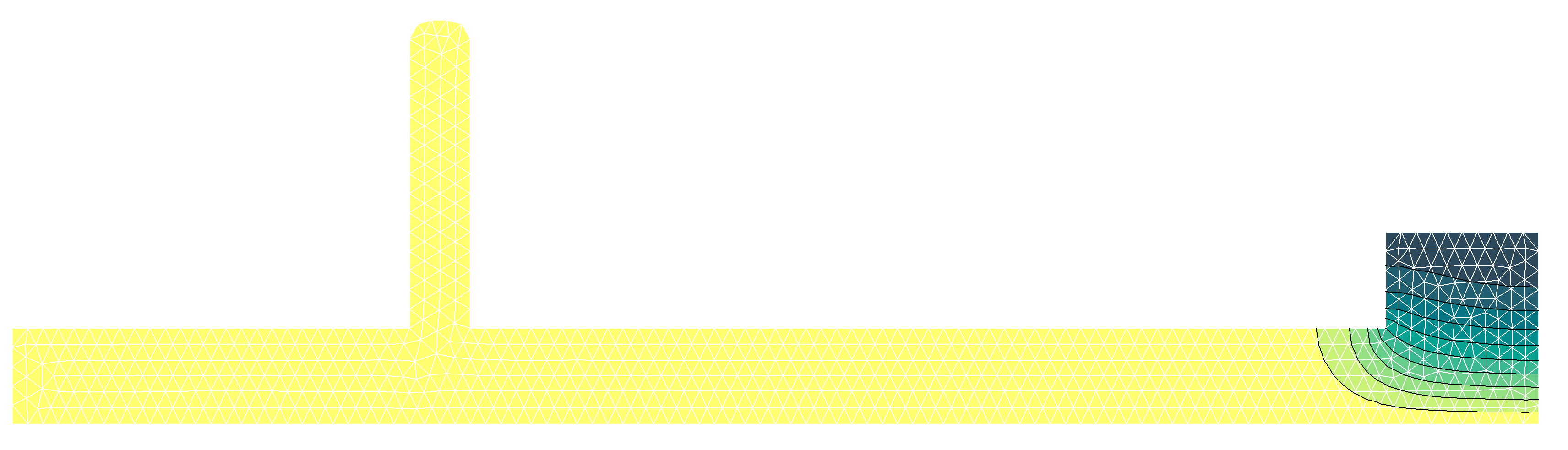}
        \centering 
        \caption{$9^{th}$ mode.}
        \label{cil_charge_dc_slot_5}
        \end{subfigure}
    \end{center}
    \caption{Cylindrical slotted charge with combustion of the bases and $D_c$=0.1m.}
    \label{response_burning_2_5}
\end{figure} 

\noindent Finally, for the geometry \ref{cil_charge_dc_slot} , the variety of mode types are shown in Figure \ref{cil_charge_dc_slot_1_l} . The first mode corresponds, as usual, to a longitudinal mode, as can be seen by the influence of the nozzle. For the second and third modes, a progressive low influence of the nozzle is seen, since in the fourth mode a transverse mode is formed in the slot region \ref{4mode_6}, meaning a very low influence of the nozzle, practically null to the stability. The stability of the fifth, sixth, and seventh modes can be regarded as the triplet of modes mentioned before. Lastly, the eighth mode is similar to the modes 5 and 6, and the ninth mode shows a high influence of the nozzle on the stability due to the formation of a transverse mode at the nozzle region. 

\subsubsection{Burning surface response function parameters} 

\begin{figure}[H]
    \includegraphics[width = 0.6\textwidth]{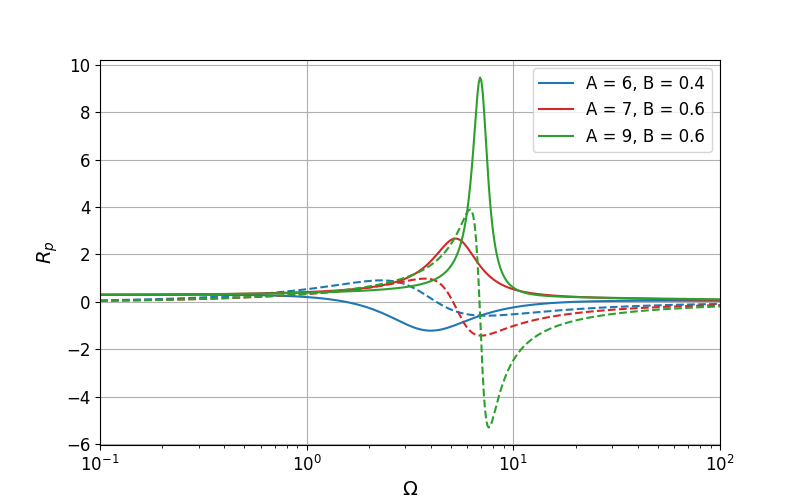}
    \centering
    \caption{Burning surface transfer function for different parameters. Bold line: Real part. Dashed Line: Imaginary part.}
    \label{A9B06_2}
\end{figure}

\begin{figure}[H]
\begin{center}
    \begin{subfigure}{8cm}
        \includegraphics[width = \textwidth]{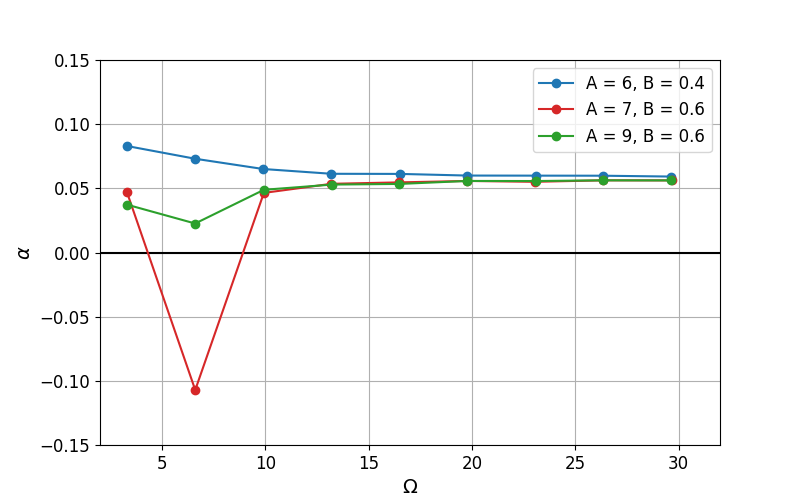}
        \centering
        \caption{Combustion chamber \ref{cil_charge}.}
        \label{cil_charge_AB_2}
    \end{subfigure}
    \hspace{1pt}
    \begin{subfigure}{8cm}
        \includegraphics[width = \textwidth]{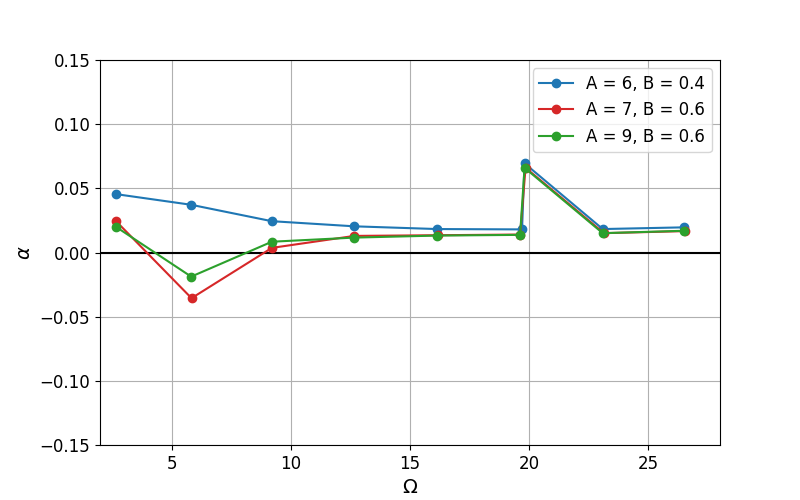}
        \centering
        \caption{Combustion chamber \ref{cil_charge_dc}.}
        \label{cil_charge_dc_AB_5}
    \end{subfigure}
    \begin{subfigure}{8cm}
        \includegraphics[width = \textwidth]{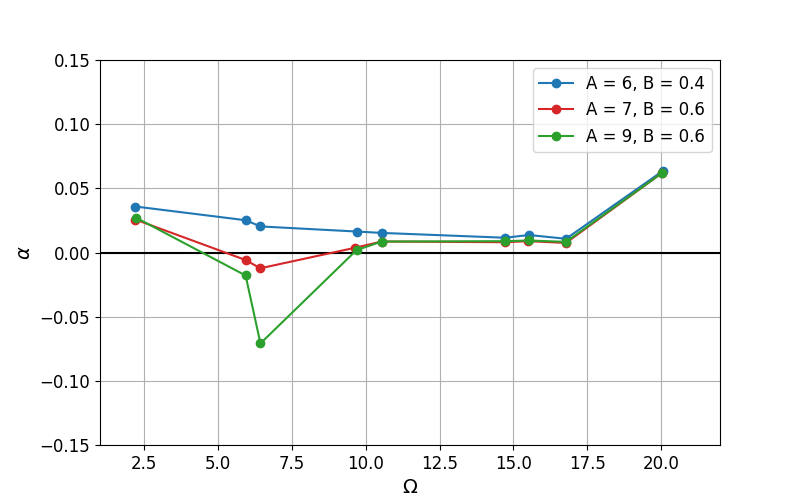}
        \centering
        \caption{Combustion chamber \ref{cil_charge_dc_slot}.}
        \label{cil_charge_dc_slot_AB_4}
        \end{subfigure}
\end{center}

    \caption{Stability value of the modes of each combustion chamber varying the parameters of the burning surface response function.}
    \label{AB_combustion_response_3}
\end{figure} 

The parameters of the burning surface response function have been varied in order to study its influence over the three different geometries with nozzle damping, characterised by the limit value of the dimensionless frequency  $\Omega_c=100$. \\

\noindent It is observed in Figure \ref{A9B06_2} that, depending on the parameters, the real part of the combustion response function changes its sign. Therefore, it can be noticed in Figure \ref{AB_combustion_response_3} that the parameter change stabilises the modes contained in the region of the major influence of the combustion response function. Nevertheless, it is seen that the instability magnitude of the response function can be increased, raising the instability of the modes whose dimensionless frequencies correspond to the frequency range of the combustion function response. \\

\noindent On the other hand, it can be concluded that the changes in the parameters of the burning response function do not significantly affect the modes that are not contained in the high influence frequency range of the burning surface transfer function.

\subsubsection{Nozzle response function parameters} 

\begin{figure}[H]
\begin{center}
    \begin{subfigure}{8cm}
        \includegraphics[width = \textwidth]{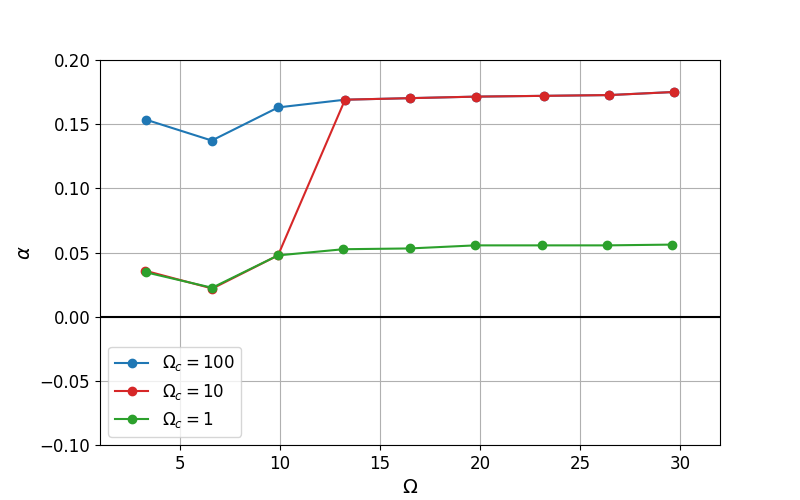}
        \centering
        \caption{Combustion chamber \ref{cil_charge}.}
        \label{cil_charge_F_3}
    \end{subfigure}
    \hspace{1pt}
    \begin{subfigure}{8cm}
        \includegraphics[width = \textwidth]{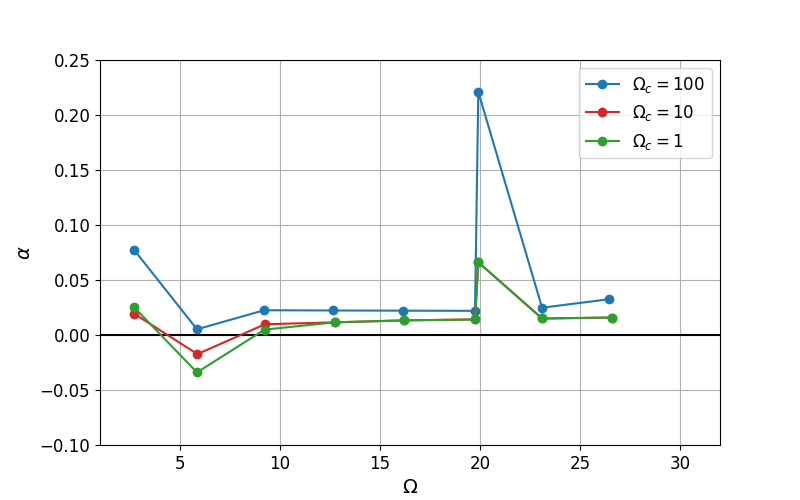}
        \centering
        \caption{Combustion chamber \ref{cil_charge_dc}.}
        \label{cil_charge_dc_F_1}
    \end{subfigure}
    \begin{subfigure}{8cm}
        \includegraphics[width = \textwidth]{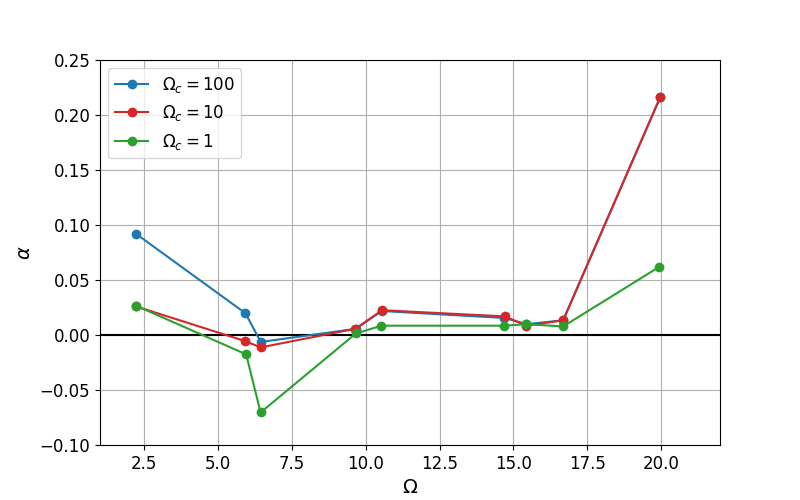}
        \centering
        \caption{Combustion chamber \ref{cil_charge_dc_slot}.}
        \label{cil_charge_dc_slot_F_2}
        \end{subfigure}
    \end{center}
    \caption{Stability value of the modes of each combustion chamber varying limit value of the dimensionless frequency of the nozzle response function.}
    \label{F_nozzle_response_2}
\end{figure}

In order to study the influence of the response function, the limit value of the dimensionless frequency has been varied for the three different geometries.\\

\noindent From the Figure \ref{cil_charge_F_3} it can be concluded that, for longitudinal modes, the effect is quite substantial due to the influence of the nozzle function response previously mentioned. Therefore, the effect of the variation of this parameter is easier to appreciate. Nevertheless, for the rest of the combustion chambers with a wider variety of modes, it is not as easy to correctly distinguish its influence, as in some of these modes, the influence of the nozzle response function has been observed to be practically null.

\subsubsection{Effect of the cavity geometry}

This section focuses on the study of the effect of the variation of the geometry of the combustion chamber, mainly to analyze the evolution of the stability value due to the increase of the advance parameter of the burning surface. To study the evolution of the stability value due to the regression of the burning surface,  the cylindrical combustion chamber \ref{cil_charge} has been used. The stability value has been studied for three combustion times. 

\begin{table}[H]
  \centering
    \begin{tabular}{c |c }
    \hline
    $t_b$ [s] & $D_c$ [m] \\
    \hline
    0    & 0.05 \\
    10     & 0.15 \\
    20     & 0.25  \\
    \hline
    \end{tabular}%
    \caption{Variation of the diameter of the cylindrical combustion chamber during combustion for $r_b$=0.01 m/s. }
  \label{tab:cil_charge_dc_slot_t1_2}%
\end{table}%

\begin{figure}[H]
    \includegraphics[width = 0.7\textwidth]{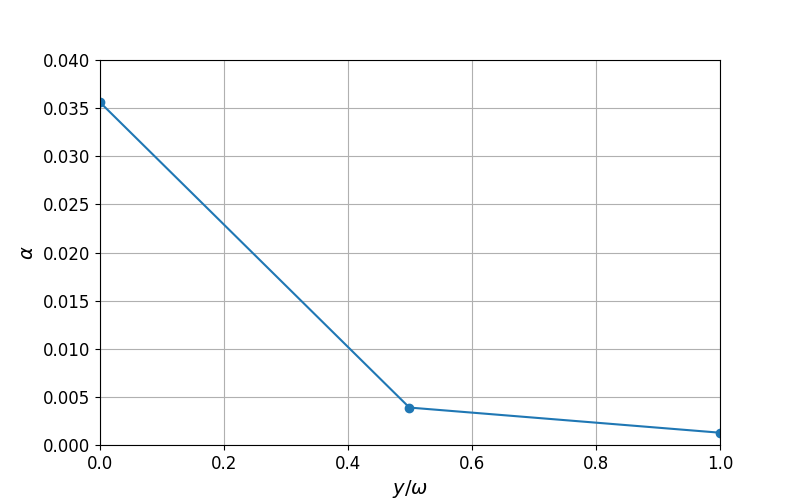}
    \centering
    \caption{Representation of the stability parameter evolution vs. the regression of the burning surface.}
    \label{tb_regress_2}
\end{figure}

\noindent As stated in the literature, the evolution of the stability margin tends to decrease due to the increase of the advance parameter \cite{Tizon2018}. Therefore, it can be seen how the volume integrals increase while the surface integrals remain constant. Thus, it will be more probable for the combustion chamber to experience combustion instabilities at the end of the combustion.

\subsubsection{Variation of the initial parameters}

This section will analyze the influence of the different propellant and motor parameters on a fixed geometry of the combustion chamber. The geometry is meshed with dimensionless longitudes in regard to the characteristic mesh longitude $L_m$. Therefore, the surface and volumetric integrals of the modal space are dimensionless regarding the previous characteristic mesh longitude. Moreover, considering that $\Bar{a}\simeq c^*$ , a characteristic mesh time can be defined as
\begin{equation}
    t_m = \frac{L_m}{c^*}
\end{equation}

\noindent On the other hand, it is possible to define a characteristic time $t_c$ and longitude $l_c$ as 
\begin{equation}
    t_c = \frac{\alpha_p}{\Dot{r}_p^2} \quad\quad l_c = c^*t_c
\end{equation}

\noindent Recalling the expression of the variable $k^*$ it can be stated that $k^*=l_c^{-1}$. Consequently, a dimensionless time can be defined by associating both characteristic times as 
\begin{equation}
    \tau = \frac{t_c}{t_m} = \frac{\alpha_pc^*}{\Dot{r}_p^2L_m}
\end{equation}

\noindent Therefore, for a fixed geometry of combustion chamber, it is possible to observe how the variation of propellant and motor parameters, which vary the value of $\tau$, influence the stability value.\\

\begin{figure}[H]
    \includegraphics[width = 0.7\textwidth]{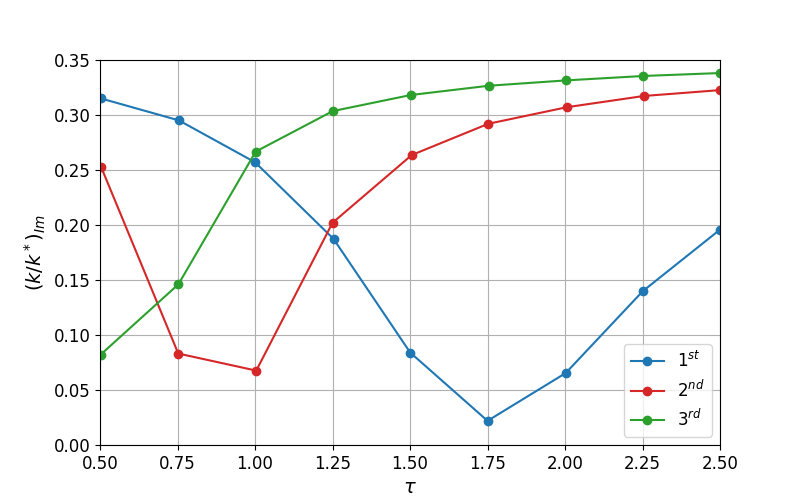}
    \centering
    \caption{Evolution of stability values of the first three modes of the combustion chamber \ref{cil_charge} versus the dimensionless characteristic time $\tau$.}
    \label{alpha_tau}
\end{figure}

\noindent The Figure \ref{alpha_tau} represents the variation of the stability value due to the variation of the dimensionless time $\tau$. It is observed that a value of $\tau$ between 1.5 and 2 unsettles the first mode. Nevertheless, it can also be extracted from the graph, that values of $\tau$ that stabilise one mode, can destabilise a different mode. This is because the delay response has an associated range of dimensionless frequencies. Therefore, the variation of the parameters of the propellant and the motor can produce the instability of some modes or the stability of another.

\section{Conclusions}
An analysis of high-frequency combustion instabilities in solid rocket motors has been carried out, focusing on the calculation of acoustic modes and frequencies within complex combustion chamber geometries. Using a known theoretical framework, the research integrates combustion dynamics with acoustic behaviour to predict chamber stability under different conditions. 

\vspace{5pt}

\noindent A key aspect of this work is the application of a numerical model based on the discretisation of the Laplacian operator using an unstructured mesh. This method, implemented by a specialised numerical algorithm, allows the accurate calculation of acoustic eigenfrequencies and modes in complex geometries. Computational examples demonstrate the effectiveness of this approach in real-world scenarios, particularly in rocket motor configurations with intricate geometries. These examples provide a practical way of integrating instability analysis into the design process. The results provide engineers with valuable data that can be used to improve chamber geometry and assess the risk of instabilities, thereby helping to optimise both structural integrity and propulsion performance.  

\vspace{5pt}

\noindent While the theoretical understanding of high frequency instabilities is well established, the implementation of unstructured mesh-based numerical models provides an effective tool for their detailed analysis. This approach enhances the ability to predict and evaluate instabilities in solid rocket motors, providing a useful resource for the design and optimisation of propulsion systems.

\section{Acknowledgments}
This study has been carried out as part of the PILUM project (Proyecto de Investigación de tecnologías para Lanzador, Ubicado en plataforma aérea, de Micro y nano satélites – PCD 2014140288) promoted by INTA (Instituto Nacional de Tecnología Aeroespacial Esteban Terradas), an autonomous agency of the Spanish public administration responsible for the aerospace and defense technologies research. It is especially appreciated the support received from Tte. Cnel. Jesús Sánchez, head of the Department of Rockets and Orthotronics at INTA-Marañosa Campus, without whose technical and material support this work would not have been possible.

%
%  bibliography
%
\printbibliography[
title={References}
]

@article{Culick2006,
author = {Culick, F. E. C. and Kuentzmann, P.},
year = {2006},
month = {12},
pages = {},
title = {{Unsteady Motions in Combustion Chambers for Propulsion Systems}},
journal = {NATO RTO-AG-AVT-039, AGARDograph}
}

@book{sutton2017rocket,
  title={{Rocket Propulsion Elements}},
  author={Sutton, G. P. and Biblarz, O.},
  year={2017},
  edition={9th edition},
  journal={John Wiley \& Sons},
}

@article{tizón2023trimpackunstructuredtriangularmesh,
      title={{Trimpack: Unstructured Triangular Mesh Generation Library}}, 
      author={J. M. Tizón and N. Becerra and D. Bercebal and C. P. Grabowsky},
      year={2023},
      eprint={2302.02795},
      journal={arXiv},
      primaryClass={cs.MS},
      url={https://arxiv.org/abs/2302.02795}, 
}

@article{Culick1968,
author = {Culick, F. E. C.},
year = {1968},
number = {No. 12},
pages = {},
title = {{A Review of Calculations for Unsteady Burning of Solid Propellant}},
volume = {Vol. 6},
journal = {AIAA Journal},
doi = {10.2514/3.4980}
}

@book{Tizon2018,
  author    = {Jenaro de Mencos, G. and Tizón Pulido, J. M.},
  title     = {{Propulsión de Misiles Tácticos}},
  year      = {2019},
  publisher = {Ed. Garceta, Madrid},
    journal = {Ed. Garceta, Madrid}
}

@article{TSIEN1952,
author = {Tsien, H. S.},
title = {{The Transfer Functions of Rocket Nozzles}},
journal = {Journal of the American Rocket Society},
volume = {Vol. 22 },
number = {No. 3},
pages = {139-143},
year = {1952},
doi = {10.2514/8.4448},
URL = {  
        https://doi.org/10.2514/8.4448
},
eprint = { 
        https://doi.org/10.2514/8.4448
}
}

@misc{Spectra,
  author       = {Y. Qiu},
  title        = {{Spectra: Sparse Eigenvalue Computation Toolkit}},
  year         = {2015--2022},
  howpublished = {{\hypersetup{urlcolor=black}\url{https://spectralib.org}}},
}

@book{Price1959,
author="Price, E. W.",
title="{Combustion Instability in Solid Propellant Rocket Motors}",
year="1959",
publisher="Springer Vienna",
pages="865--874",
}

@article{Poinsot2017,
   author = {T. Poinsot},
   doi = {10.1016/j.proci.2016.05.007ï},
   issn = {1540-7489},
   issue = {1},
   journal = {Proceedings of the Combustion Institute},
   pages = {1-28},
   publisher = {Elsevier},
   title = {Prediction and control of combustion instabilities in real engines},
   volume = {Vol. 36},
   url = {https://hal.archives-ouvertes.fr/hal-01502419},
   year = {2017},
}

@article{Yang1995,
  author    = {V. Yang and W. E. Anderson},
  title     = {{Liquid Rocket Engine Combustion Instability}},
  series    = {Progress in Astronautics and Aeronautics},
  volume    = {Vol. 169},
  year      = {1995},
  journal = {Progress in Astronautics and Aeronautics, AIAA},
  address   = {Washington, D.C.}
}

@article{Huang2009,
  author    = {Y. Huang and V. Yang},
  title     = {{Dynamics and Stability of Lean-Premixed Swirl-Stabilized Combustion}},
  journal   = {Progress in Energy and Combustion Science},
  volume    = {Vol. 35},
  number    = {No. 4},
  pages     = {293--364},
  year      = {2009},
  doi       = {10.1016/j.pecs.2009.01.002}
}

@book{Poinsot2005,
   author = {Poinsot, T. and Veynante, D.},
   title = {{Theoretical and Numerical Combustion}},
   edition = {2nd edition},
   year = {2005},
   journal = {Edwards},
   pages = {473}
}

@article{Flandro1995,
author = {Flandro, G. A.},
title = {Effects of vorticity on rocket combustion stability},
journal = {Journal of Propulsion and Power},
volume = {Vol. 11},
number = {No. 4},
pages = {607-625},
year = {1995},
doi = {10.2514/3.23887},
URL = { 
        https://doi.org/10.2514/3.23887
},
eprint = { 
        https://doi.org/10.2514/3.23887
}
}

@article{Dowling2003,
author = {Dowling, A. and Stow, S.},
year = {2003},
month = {No. 09},
pages = {751-764},
title = {{Acoustic Analysis of Gas Turbine Combustors}},
volume = {Vol. 19},
journal = {Journal of Propulsion and Power},
doi = {10.2514/2.6192}
}

@article{EMELYANOV2017161,
title = {Pressure oscillations and instability of working processes in the combustion chambers of solid rocket motors},
journal = {Acta Astronautica},
volume = {Vol. 135},
pages = {161-171},
year = {2017},
issn = {0094-5765},
doi = {https://doi.org/10.1016/j.actaastro.2016.09.029},
url = {https://www.sciencedirect.com/science/article/pii/S0094576516308153},
author = {V.N. Emelyanov and I.V. Teterina and K.N. Volkov and A.U. Garkushev},
}

@article{VO201812,
title = {Moving boundary modeling for solid propellant combustion},
journal = {Combustion and Flame},
volume = {Vol. 189},
pages = {12-23},
year = {2018},
issn = {0010-2180},
doi = {https://doi.org/10.1016/j.combustflame.2017.09.040},
url = {https://www.sciencedirect.com/science/article/pii/S0010218017303784},
author = {Nguyen Dat Vo and Min Young Jung and Dong Hoon Oh and Jung Soo Park and Il Moon and Min Oh},
}

@misc{DISLIN,
  author       = {Helmut Michels},
  title        = {DISLIN Data Plotting Software},
  year         = {1986-2024},
  howpublished = {{\hypersetup{urlcolor=black}\url{http://www.dislin.de}}},
}

@article{marble1977acoustic,
  author = {Marble, F. E. and Candel, S. M.},
  title = {Acoustic disturbance from gas non-uniformities convected through a nozzle},
  journal = {Journal of Fluid Mechanics},
  volume = {Vol. 55},
  number = {No. 2},
  pages = {225--243},
  year = {1977},
  doi = {10.1016/0022-460X(77)90596-X}
}

\end{document}